%% file: HybridMultiObserverSecondVersion.tex
\documentclass[journal,twoside,web]{ieeecolor}

\usepackage{generic}
\usepackage{amsmath,amssymb,amsfonts}
\usepackage{algorithmic}
\usepackage{graphicx}
\usepackage{textcomp}
\def\BibTeX{{\rm B\kern-.05em{\sc i\kern-.025em b}\kern-.08em
    T\kern-.1667em\lower.7ex\hbox{E}\kern-.125emX}}

\usepackage{graphicx,amstext,amsmath,amssymb,color,psfrag,mathabx,booktabs}
\usepackage[colorlinks=true, letterpaper,    
urlcolor=black,     
citecolor=blue,
menucolor=black,    
linkcolor=black,    
pagecolor=black,    
breaklinks=true,
]{hyperref}

\usepackage{amsthm}

\usepackage{comment}

\include{commands}

\usepackage{array}

\usepackage{mathtools}

\usepackage[noadjust]{cite}

\let\labelindent\relax
\usepackage{enumitem}
\usepackage{tikz}
\usetikzlibrary{shapes,arrows,calc}
\definecolor{MyGreen}{RGB}{50,140,80}

\usepackage[export]{adjustbox}
\usepackage{ circuitikz }
\usetikzlibrary{arrows}

\newcommand{\overbar}[1]{\mkern 1.5mu\overline{\mkern-1.5mu#1\mkern-1.5mu}\mkern 1.5mu}

\usetikzlibrary{backgrounds}

\begin{document}
\title{
	Hybrid multi-observer for improving estimation performance 
}

\author{E. Petri, 	R. Postoyan, D. Astolfi, D. Ne\v{s}i\'c and V. Andrieu
	\thanks{This work was funded by Lorraine Universit\'e d'Excellence LUE, the France Australia collaboration project IRP-ARS CNRS and the Australian Research Council under the Discovery Project DP200101303.  }
	\thanks{
		E. Petri is with the Department of Mechanical Engineering, Eindhoven University of Technology, The Netherlands.   
		(e-mail:e.petri@tue.nl).}
			\thanks{
			 R. Postoyan is with the Universit\'e de Lorraine, CNRS, CRAN, F-54000 Nancy, France.   
			(e-mail: romain.postoyan@univ-lorraine.fr).}
	\thanks{D. Astolfi and V. Andrieu are with  
		Universit\'e Claude Bernard Lyon 1, CNRS, LAGEPP UMR 5007, F-69100,
		Villeurbanne, France. (e-mail:daniele.astolfi@univ-lyon1.fr, vincent.andrieu@univ-lyon1.fr).}
	\thanks{ D. Ne\v{s}i\'c is with the Department of Electrical and Electronic Engineering, The University of Melbourne, Parkville, 3010 Victoria Australia.
		(e-mail:dnesic@unimelb.edu.au). }}
\maketitle
	\begin{abstract}
	Various methods are nowadays available to design observers for broad classes of systems, where the primary focus is on establishing the convergence of the estimated states. Nevertheless, the question of the tuning of the observer to achieve satisfactory estimation performance remains largely open. In this context, we present a general design framework for the online tuning of the observer gains. 
	Our starting point is a robust nominal observer designed for a general nonlinear system, for which an input-to-state stability property can be established. Our goal is then to improve the performance of this nominal observer. We present for this purpose a new hybrid multi-observer scheme, whose flexibility can be exploited to enforce various desirable properties,  e.g., fast convergence and good sensitivity to measurement noise. 
	We prove that an input-to-state stability property also holds for the proposed scheme and, importantly, we ensure that the estimation performance in terms of a quadratic cost is (strictly) improved. 
	We illustrate the efficiency of the approach in improving the performance of given nominal observers in two numerical examples (Van der Pol oscillator and Lithium-Ion (Li-Ion) battery model). 	
\end{abstract}

\section{Introduction}\label{Introduction}
State estimation of dynamical systems is a central theme in control theory, whereby an observer is designed to estimate the unmeasured system states exploiting the knowledge of the system model and input and output measurements. 
Many techniques are available in the literature for the observer design of linear and nonlinear systems, see \cite{bernard2022observer} and the references therein. 
The vast majority of these works focuses on designing the observer so that the origin of the associated estimation error system is (robustly) asymptotically stable. 
A critical and largely open question is how to tune the observer to obtain desirable performance (e.g. convergence speed and overshoot in the transient response) in the presence of model uncertainties, measurement noise and disturbances. This question obviously also arises in the context of control.

One of the challenges in observer tuning is that there exist different trade-offs between the desired properties. 
%
Indeed, a standard approach to observer design consists in using a copy of the plant (in some coordinates that may be different from the original ones) and then adding a correction term,  often denoted as the output injection term. This term is designed by multiplying the difference between the measured output and the estimated one by a (possibly nonlinear) gain, whose tuning produces different estimation performance. 
Typically, the output injection term with small gains produces an observer robust to measurement noise, but its convergence is very slow. On the contrary, an observer with a large gain usually has a fast convergence, but is more sensitive to noise. 
An answer to the question on how to tune the observer gain in the special context of linear systems affected by additive Gaussian noise impacting the dynamics and the output is the celebrated Kalman filter \cite{kalman1961new}. 
For general nonlinear systems and noise/disturbances, optimal observer design with global stability guarantees is notoriously hard.
	%
For instance, in the context of nonlinear systems, optimal state estimation requires solving challenging partial differential equations \cite{helton1999extending}. In this context, an alternative consists in aiming at \textit{improving} 
the estimation performance of a given observer. 
To the best of the authors' knowledge, existing works in this direction either concentrate on specific classes of systems (see e.g., \cite{li2015finite, rios2018hybrid, alessandri2022hysteresis,popov2017adaptive} for linear systems or e.g., \cite{astolfi2015high, astolfi2018low, esfandiari2019bank, bernat2015multi, ahrens2009high, farza2021improved} in the context of high-gain observers), specific observers e.g., \cite{reif1998ekf, bonnabel2014contraction, dai2022q} or specific properties like robustness to measurement noise (see, e.g.,  \cite{astolfi2021stubborn, battilotti2021performance}) or the reduction of the undesired effect of the peaking phenomenon (see e.g., \cite{astolfi2021constrained} for a general approach and \cite{astolfi2018low,bernat2015multi, farza2021improved} for specific solutions in the context of high-gain observer).
An exception is \cite{astolfi2019uniting}, where two observers designed for a general nonlinear system are ``united'' to exploit the good properties of each. However, the design in \cite{astolfi2019uniting} is not always easy to implement as it requires knowledge of various properties of the observers (basin of attraction, ultimate bound), which may be difficult to obtain. Alternative methods proposed in the literature are particle filters and unscented Kalman filters, see e.g., \cite{sarkka2007unscented}, which combine observers to obtain good estimation performance, but are not endowed with robust stability guarantees in general.
Recently, in \cite{schiller2022suboptimal, schiller2022simple} suboptimal moving horizon estimation schemes have been proposed for general discrete-time nonlinear systems, where the performance of a given auxiliary observer is improved.

In this context, we present a flexible and general observer design methodology based on supervisory multi-observer ideas that can be used to address various trade-offs between robustness to modeling errors and measurement noise, and convergence speed.
A multi-observer consists of a bank of observers that run in parallel. It has been used in the literature in a range of different contexts, including improvement of the sensitivity to measurement noise and/or reduction of the undesired overshoot during the transient (e.g., \cite{esfandiari2019bank, bernat2015multi} for high-gain observer and \cite{peralez2022neural} for KKL observer), joint state-parameter estimation (e.g., \cite{chong2015parameter, aguiar2008identification, aguiar2007convergence, meijer2021joint, cuevas2020multi}) or distributed observers (e.g., \cite{han2018simple, park2016design, wang2022hybrid}).
%
%
%
In this paper, we propose a new problem formulation that we believe has not yet been addressed in the literature. 
Our starting point is the knowledge of a nominal observer, which ensures that the associated state estimation error system satisfies an input-to-state stability property with respect to measurement noise and disturbances. Various methods from the literature can be used to design the nominal observer, see  \cite{astolfi2021stubborn, shim2015nonlinear} and the references therein. Then, we
construct a multi-observer, composed of the nominal observer and additional dynamical systems, all together called \emph{modes}, that have the same structure as the nominal observer, but different gains. It is important to emphasize that  the number of modes and the associated gains can 
be freely assigned (no specific stability/convergence property is required). 
%
Because the gains are different, each mode exhibits different properties in terms of speed of convergence and robustness to measurement noise. 
%
	%
	The latter point is relevant as we can heuristically design the extra modes for the application at hand without having to worry about the proof of their convergence. For instance, we can select one of these modes to have a zero gain, which is advantageous when measurement noise is an issue, as we illustrate on simulations. We will provide more general guidelines for the design of the extra modes.
We run all modes in parallel and we evaluate their estimation performance in terms of a quadratic cost using monitoring variables, inspired by supervisory control and estimation techniques, see e.g., \cite{chong2015parameter, cuevas2020multi, hespanha2003hysteresis, morse1996supervisory, hespanha2001multiple, peralez2020data}. 
Based on these running costs (i.e., monitoring variables), we design a switching rule that selects, at any time instant, the mode which is providing the best performance. When a new mode is selected, the other ones may reset or not their current state estimate (and  their monitoring variable) to it.

We model the overall system as a hybrid system using the formalism of \cite{goebel2012hybrid}. 
We prove that the proposed hybrid estimation scheme satisfies an input-to-state stability property with respect to disturbance and measurement noise. Note that such a property is not obvious as we do not require any convergence guarantee on the additional modes, but only for the nominal one.
We also guarantee the existence of a (semiglobal uniform) average dwell-time thereby ruling out Zeno phenomenon.
The performance of the proposed hybrid multi-observer in terms of the cost associated with the designed monitoring variables is guaranteed to be, at least, as good as the performance of the nominal observer by design. Moreover, we provide extra conditions under which the proposed hybrid multi-observer produces a strict performance improvement compared to the nominal one in terms of an integral cost. 	
The efficiency of the proposed technique is illustrated on two numerical examples. In the first one, the proposed approach is used to improve the estimation performance of a high-gain observer used to estimate the state of a Van der Pol oscillator. In the second example, we consider an equivalent circuit model for an Li-Ion battery, whose state is estimated with an observer designed using a polytopic approach and we implement the hybrid estimation scheme to improve its performance. Another example is given in \cite{petri2023state}, where a previous version of the hybrid estimation scheme \cite{petri2022towards} is applied for the state estimation of a more advanced Li-Ion battery model, namely a single-particle electrochemical model. 


Compared to the preliminary version of this work \cite{petri2022towards}, the hybrid estimation scheme is modified. As a result, we can now guarantee the existence of an average dwell-time, which excludes the Zeno phenomenon. 
 Moreover, we analyze the performance of the proposed hybrid scheme, instead of solely relying on simulations, and we provide conditions under which strict performance improvement occurs as mentioned above. 
\noindent\textbf{Notation and Preliminaries.}
$\R$ stands for the set of real numbers, $\Rlo:= [0, +\infty)$, 
 $\Z$ is the set of integers, $\Zo:= \{0,1,2,...\}$ and $\Zp:= \{1,2,...\}$. For a vector $x \in \R^n$, $|x|$ denotes its Euclidean norm. For a matrix $A \in \R^{n  \times m}, \norm{A}$ stands for its 2-induced norm. For a signal $v: \R_{\geq0} \to \R^{n_v}$ with $n_v \in \Zp$, and $t_1,t_2\in\R_{\geq 0}\cup\{\infty\}$ with $t_1\leq t_2$,  $\norm{v}_{[t_1, t_2]}:= \textnormal{ess.} \sup_{t \in [t_1, t_2]} |v(t)|$. 
Given a real, symmetric matrix $P$, its maximum and minimum eigenvalues are denoted by $\lambda_{\max}(P)$ and $\lambda_{\min}(P)$ respectively. The notation $I_N$ stands for the identity matrix of dimension $N \in \Zp$, while $0_{N \times M}$ stands for the null matrix of dimension $N \times M$, with $N, M \in \Zp$. We use $\mathbb{S}^{N}_{> 0}$ ($\mathbb{S}^{N}_{\geq 0}$) to denote the set of real symmetric positive definite (semidefinite) matrices of dimension $N$. 
The notation $\delta_{i,j}$, with $i,j \in \Z_{>0}$ denotes the Kronecker delta defined as  $\delta_{i,j} = 0$ if $i \neq j$ and $\delta_{i,j} = 1$ if $i = j$.
%
A continuous function $\alpha: [0, \infty) \rightarrow [0, \infty)$ is of class $\K$ if $\alpha(0) = 0$ and it is strictly increasing. Moreover, $\alpha$ is of class $\Kinf$ if $\alpha \in \K$ and $\lim_{r \rightarrow \infty} \alpha(r) = \infty$. A continuous function $\beta: [0,\infty) \times [0, \infty) \rightarrow [0, \infty)$ is of class $\KL$ if, for any fixed $s \in \R_{\geq0}$, $\beta(\cdot,s) \in \K$  and, for each fixed $r \in \R_{\geq0}$, $\beta (r,\cdot)$ is non-increasing and satisfies $\lim_{s \rightarrow \infty} \beta(r,s) = 0$.
%
%
%
Given a function $f: \mathcal{S}_1\to\mathcal{S}_2$ with sets $\mathcal{S}_1$, $\mathcal{S}_2$, $\dom f:=\{z\in\mathcal{S}_1 : f(z)\neq\emptyset\}$.
%
Based on the formalism of \cite{goebel2012hybrid}, we will model the proposed estimation scheme together with the plant as a hybrid system with inputs of the form
\begin{equation}
	\mathcal{H} \;:\; \left\{
	\begin{array}{rcll}
		\dot x &=& F(x,u), & \quad x \in \mathcal{C}, 
		\\
		x^+ & \in & G(x,u),  &\quad x \in \mathcal{D},
	\end{array}
	\right.
	\label{eq:hybridSystemNotation}
\end{equation}
where 
$\mathcal{C}\subseteq \R^{n_x} $ is the flow set, 
$\mathcal{D}\subseteq \R^{n_x}$ is the jump set,
$F$ is the flow map and $G$ is the jump map. 
We consider hybrid time domains as defined in \cite{goebel2012hybrid}.  The notation $(t,j) \geq (t^\star, j^{\star'})$ means that $t \geq t^\star$ and $j \geq  j^{\star'}$,  where $(t,j),(t',j')\in\R_{\geq 0}\times\Zo$.
We use the notion of solution for system \eqref{eq:hybridSystemNotation} as given in \cite[Definition 4]{heemels2021hybrid}.
Given a set $\mathcal{U} \subseteq \R^{n_u}$, $\mathcal{L}_{\mathcal{U}}$ is the set of all functions from $\R_{\geq0}$ to $\mathcal{U}$ that are Lebesgue measurable and locally essentially bounded. 
%
Given a set $\mathcal{C} \subseteq \R^{n_x}$, the tangent cone to the set $\mathcal{C}$ at a point $x \in \R^{n_x}$, denoted $T_\mathcal{C}(x)$, is the set of all vectors $v \in \R^{n_x}$ for which there exist $x_i \in \mathcal{C}$, $\tau_i > 0$, with $x_i \to x$, $\tau_i \to 0$, and $v = \lim_{i \to \infty} \frac{x_i - x}{\tau_i}$.
Given a solution $q$ for system \eqref{eq:hybridSystemNotation} and its hybrid time domain $\dom q$,  $I^j := \{t:(t,j) \in \dom q\}$ is the flow interval for each $j \in \Zo$ and $\textnormal{int} \  I^j$ denotes its interior. 
Finally, we use $U^\circ(x;v):= \limsup_{h \to 0^+, y \to x}\frac{U(y+hv) - U(y)}{h}$ to denote the Clarke generalized directional derivative of a Lipschitz function $U$ at $x$ in the direction $v$ 
\cite{clarke1990optimization}.
%

\section{Problem statement}\label{ProblemStatement}
The aim of this work is to improve the estimation performance of a given nonlinear nominal observer by exploiting a novel hybrid estimation scheme that is presented in the next section. 
%
%
We consider the plant model
\begin{equation}
	\begin{aligned}
		\dot x  &=  f_p(x,u,v)\\
		y  &=  h(x, w),
	\end{aligned}
	\label{eq:system}
\end{equation}
where $x \in \R^{n_x}$ is the state to be estimated, $u \in \R^{n_u}$ is the measured input, 
$y \in \R^{n_y}$ is the measured output, $v \in \R^{n_v}$ is an unknown disturbance input and $w \in \R^{n_w}$ is an unknown measurement noise, with $n_x, n_y \in \Zp$ and $n_u, n_v, n_w \in \Zo$. The input signal $u: \R_{\geq 0} \rightarrow \R^{n_u}$, the unknown disturbance input $v:\R_{\geq 0} \rightarrow \R^{n_v}$ and the measurement noise $w:\R_{\geq 0} \rightarrow \R^{n_w}$ are such that $u \in \mathcal{L_U}$, $v \in \mathcal{L_V}$ and $w \in \mathcal{L_W}$ for closed sets $\mathcal{U} \subseteq \R^{n_u}$,  $\mathcal{V} \subseteq \R^{n_v}$ and $\mathcal{W} \subseteq \R^{n_w}$. 

We consider a so-called nominal observer for system~\eqref{eq:system} of the form
\begin{equation}
	\begin{aligned}
		\dot{\hat{x}}_{1}  &=  f_o(\hat{x}_{1},u, L_{1} (y-\hat{y}_{1}))\\
		\hat{y}_{1}  &=  h(\hat{x}_{1}, 0),
	\end{aligned}
	\label{eq:observerNominal}
\end{equation}
where $\hat{x}_{1}\in \R^{n_x}$ is the state estimate, $\hat{y}_{1} \in \R^{n_y}$ is the output estimate and $L_{1} \in \R^{n_{L_1} \times{n_y}}$ is the observer output injection gain with $n_{L_1} \in \Zp$.
We define the estimation error as $e_{1} := x -\hat{x}_{1} \in \R^{n_x}$ and introduce a \textit{perturbed} version of the error dynamics, following from \eqref{eq:system} and \eqref{eq:observerNominal}, as 
\begin{equation}
	\begin{aligned}
		\dot e_1 &= f_p(x,u,v)- f_o(\hat{x}_{1},u, L_{1}(y - \hat{y}_{1}) +d )\\
		&=:\tilde f(e_1,x,u, v, w, d)
	\end{aligned}
	\label{eq:PerturbedNominalEstimationError}
\end{equation}
where $d \in \R^{n_{L_1}}$ represents an additive perturbation on the output injection term $L_1(y-\hat y_1)$. 
We assume that observer \eqref{eq:observerNominal} is designed such that 
system \eqref{eq:PerturbedNominalEstimationError} is input-to-state stable with respect to $v$, $w$ and $d$, uniformly in $u$ and $x$, as formalized next. 
%
\begin{ass}
	There exist $\underline{\alpha}$,$\overline{\alpha},  \psi_1, \psi_2 \in \Kinf$, 
	$\alpha \in \R_{> 0}$, $\gamma\in \R_{\geq 0}$
	and  $V: \R^{n_x} \rightarrow \R_{\geq0}$ continuously differentiable, such that for all $x \in \R^{n_x}$, $e_{1} \in \R^{n_x}$, $d \in \R^{n_{L_1}}$, $u \in \mathcal{U}$,  $v \in  \mathcal{V}$, $w \in \mathcal{W}$, 
	\begin{equation}
		\underline{\alpha}(|e_{1}|) \leq V(e_{1}) \leq \overline{\alpha}(|e_{1}|)	
		\label{eq:NominalAssumptionSandwichBound}	
	\end{equation}
	\begin{equation}
		\begin{array}{r}
			\hspace{-0.5em} \left\langle \nabla V(e_{1}), \tilde f(e_1, x,u, v, w, d) \right\rangle  
			\leq 
			-\alpha V(e_{1}) + \psi_1(|v|) \\[0.5em] 
			+ \psi_2(|w|)  + \gamma|d|^2.
		\end{array}
		\label{eq:NominalAssumptionDerivative}
	\end{equation}	
	\label{NominalAssumption}
\end{ass}
A large number of observers in the literature have the form of \eqref{eq:observerNominal} and satisfy Assumption~\ref{NominalAssumption}, possibly after a change of coordinates, see \cite{astolfi2021constrained, astolfi2021stubborn, shim2015nonlinear} and the references therein for more details. Examples are also provided in Section~\ref{Example}. 
Assumption~\ref{NominalAssumption} implies that 
there exist $\beta \in \KL$ and $\rho \in \Kinf$ such that, 
for any $u \in\mathcal{L}_{\mathcal{U}}$, $v \in \mathcal{L_V}$, $w \in \mathcal{L_W}$ and $d \in \mathcal{L}_{\R^{n_{L_1}}}$, any corresponding solution $(x, e_1)$ to systems \eqref{eq:system} and \eqref{eq:PerturbedNominalEstimationError} verifies, for all $t\in \dom (x, e_1)$, 
\begin{equation}
	|e_{1} (t)| \leq \beta(|e_{1}(0)|, t) + \rho(\norm{v}_{[0,t]} +\norm{w}_{[0,t]} + \norm{d}_{[0,t]}).
	\label{eq:ISSnominalConverging}
\end{equation}
Inequality
\eqref{eq:ISSnominalConverging} provides a desirable robust stability property of the estimation error associated with observer \eqref{eq:observerNominal}. However, this property may not be satisfactory in terms of performance, like convergence speed and noise/disturbance rejection. 
%
%
To tune $L_1$ to obtain desirable performance properties is highly challenging in general and even impossible in some cases  when the desired properties are conflicting like high convergence speed and efficient noise rejection, see e.g., \cite{seron2012fundamental}.
To address this challenge, we propose a hybrid redesign of observer \eqref{eq:observerNominal}, which aims at improving its performance, 
in a sense made precise in the following, while preserving an input-to-state stability property for the obtained estimation error system.
\begin{rem}
%
		The results of the paper apply \emph{mutatis mutandis} to the case where Assumption \ref{NominalAssumption} holds semiglobally or when the Lyapunov function $V$ depends on both $x$ and $e_1$, which allow to cover an even broader class of observers \cite[Section~V]{bernard2022observer}.
		We do not consider these in this paper to not over-complicate the exposition and to not blur the main message of the work.
\end{rem}\label{Rem:ExtensionsToOtherObservers}

In the following we also need the next technical assumption on the output map $h$ in \eqref{eq:system}.
\begin{ass}
	There exist $\delta_1, \delta_2 \in \R_{> 0}$ such that for all $x, x' \in \R^{n_x}$, $w, w' \in \mathcal{W}$,  
	\begin{equation}
		|h(x,w) - h(x', w')|^2 \leq \delta_1 V(x-x') + \delta_2|w-w'|^2,
		\label{eq:equationAssumptionOutputMap}
	\end{equation}
	where $V$ comes from Assumption~\ref{NominalAssumption}. 
	\label{ASS:ass1}
\end{ass}
Assumption~\ref{ASS:ass1} holds in the common case where $V$ in Assumption~\ref{NominalAssumption} is quadratic and $h$ is globally Lipschitz. Indeed, in this case, 
$V(x-x'):= (x-x')^\top P(x-x')$, with $P \in \R^{n_x \times n_x}$ symmetric, positive definite, and 
$|h(x,w) - h(x',w')| \leq K|(x-x', w-w')|$ for any $x, x' \in \R^{n_x}$, $w, w' \in \mathcal{W}$ and some $K \in \R_{\geq 0}$, then \eqref{eq:equationAssumptionOutputMap} holds with $ \delta_1 = \frac{K^2}{\lambda_{\min}(P)}$ and $ \delta_2 = K^2$. Note that $h$ globally Lispchitz covers the common case where  $h(x,w)=Cx+Dw$ with $C\in\R^{n_y \times n_x}$ and $D\in \R^{n_y \times n_w}$.
%

%

\section{Hybrid estimation scheme}\label{HybridEstimationScheme}
\begin{figure*}[]
	\begin{center}
		\tikzstyle{blockB} = [draw, fill=blue!20, rectangle, 
		minimum height=3.5em, minimum width=7em]  
		\tikzstyle{blockG} = [draw, fill=MyGreen!20, rectangle, 
		minimum height=3.5em, minimum width=8.5em]
		\tikzstyle{blockR} = [draw, fill=red!40, rectangle, 
		minimum height=2em, minimum width=3em]
		\tikzstyle{blockO} = [draw,minimum height=1.5em, fill=orange!20, minimum width=4em]
		\tikzstyle{input} = [coordinate]
		\tikzstyle{blockW} = [draw,minimum height=3.5em, fill=white!20, minimum width=6em]
		\tikzstyle{blockWplant} = [draw,minimum height=1.5em, fill=white!20, minimum width=4em]
		\tikzstyle{input1} = [coordinate]
		\tikzstyle{blockCircle} = [draw, circle]
		\tikzstyle{sum} = [draw, circle, minimum size=.3cm]
		\tikzstyle{blockSensor} = [draw, fill=white!20, draw= blue!80, line width= 0.8mm, minimum height=10em, minimum width=13em]
		\tikzstyle{blockDOT} = [draw,minimum height=8em, fill=white!20, minimum width=14em, dashed]
		\tikzstyle{blockMuxGreen} = [draw, minimum height=19em, fill=MyGreen!30, line width=0.1mm]
		\tikzstyle{blockMuxBlueBig} = [draw, minimum height=19em, fill=blue!30, line width=0.1mm]
		\tikzstyle{blockMuxBlueSmall} = [draw,minimum height=6.5em, fill=blue!30, minimum width=0.8mm]
		
		\begin{tikzpicture}[auto, node distance=2cm,>=latex , scale=0.58,transform shape]  
			
			\node [input, name=input] {};
			\node [input, right of= input, node distance=0.6cm] (inputLine) {};
			\node [blockO, right of=input, node distance=1.7cm] (plant) {   
				$\begin{array}{c}
					\textbf{Plant} \\ (x)
				\end{array}$
			};
			
			\draw [draw,->] (input) -- node {} (plant);
			\draw [draw,-] (input) -- node [pos=0.3]{\large $u$} (plant);
			\node [input, above of= plant, node distance=1.3cm] (disturbance) {};
			\draw [draw,->] (disturbance) -- node [pos=0.5]{\large $v$, $w$} (plant);
			
			\node [input, right of= plant, node distance=2cm] (output) {};
			\draw [draw,-] (plant) -- node  [pos=0.5]{\large $y$} (output);
			\node [input, above of= output, node distance=3cm] (NominalObserverLine) {};
			\draw [draw,-] (output) -- (NominalObserverLine);
			\node [blockB, right of=NominalObserverLine, node distance=2.5cm] (NominalObserver) {
				$\begin{array}{c}
					\textbf{Nominal Observer}\\
					\textbf{Mode $1$}
				\end{array}$
			};
			\draw [draw,->] (NominalObserverLine) -- (NominalObserver);
			
			\node [input, above of= output, node distance=1cm] (Observer2Line) {};
			\node [blockB, right of=Observer2Line, node distance=2.5cm] (Observer2) {
				\textbf{Mode $ {2}$}
			};
			\draw [draw,->] (Observer2Line) -- (Observer2);
			
			\node [input, below of= output, node distance=3cm] (ObserverN1Line){}; 
			\draw [draw,-] (output) -- (ObserverN1Line);
			\node [blockB, right of=ObserverN1Line, node distance=2.5cm] (ObserverN1) {
				\textbf{Mode $ {N+1}$}
			};
			\draw [draw,->] (ObserverN1Line) -- (ObserverN1);
			\node at ($(Observer2)!.45!(ObserverN1)$) {\large \textcolor{blue!80}{\vdots}};
			
			\node [input, above of= NominalObserver, right = 1.55cm, node distance=0.25cm] (StateEstimate1line) {};
			\node [input, above of= NominalObserver, left = 1.70cm, node distance=0.25cm] (NominalObserverInput) {};
			\node [input, left of= NominalObserverInput, node distance=1cm] (NominalObserverInputStart) {};
			\node [input, above of= inputLine, node distance=3.25cm] (NominalObserverInputLeft) {};
			\draw [draw,-](inputLine) -- node  {}(NominalObserverInputLeft);
			\draw [draw,-](NominalObserverInputLeft) -- node  {}(NominalObserverInputStart);
			\draw [draw,->](NominalObserverInputStart) -- node  [pos=0.5]{\large \textcolor{black}{}} (NominalObserverInput);
			\node [input, right of= StateEstimate1line, node distance=0.8cm] (StateEstimate1) {};
			\node [input, right of= StateEstimate1line, node distance=0.7cm] (StateEstimate1Arrow) {};
			\node [input, left of= StateEstimate1Arrow, node distance=0.54cm] (StateEstimate1ArrowStart) {};
			\draw [draw,->, blue!80](StateEstimate1ArrowStart) -- node  [pos=0.5]{\large \textcolor{black}{$\hat x_1$}} (StateEstimate1Arrow);
			\node [input, below of= NominalObserver, right = 1.7cm, node distance=0.25cm] (MS1lineAbove) {};
			\node [input, below of= MS1lineAbove, node distance=0.25cm] (MS1line) {};
			\node [blockG, right of=MS1line, node distance=3.35cm] (MonitoringSignal1) {
				$\begin{array}{c}
					\textbf{Monitoring} \\ \textbf{variable} \ 1
				\end{array}$
				
			};
			\node [input, above of= MonitoringSignal1, left = 1.49cm, node distance=0.25cm] (MonitoringSignal1Above) {};
			
			\node [input, right of= MS1lineAbove, node distance=0.45cm] (MS1lineAboveArcStart) {};
			\node [input, right of= MS1lineAboveArcStart, node distance=0.4cm] (MS1lineAboveArcEnd) {};
			\draw [draw,-] (MS1lineAbove) -- node {} (MS1lineAboveArcStart);
			\draw [draw,->] (MS1lineAboveArcEnd) -- node  [pos=0.5]{\large $\hat y_1$} (MonitoringSignal1Above);

			\node [input, below of= MonitoringSignal1, left = 1.49cm, node distance=0.25cm] (MonitoringSignal1Below) {};
			\node [input, below of= NominalObserverLine, node distance=0.75cm] (Output1) {};
			
			\node [input, right of= Output1, node distance=4.65cm] (Output1ArcStart) {};
			\node [input, right of= Output1ArcStart, node distance=0.4cm] (Output1ArcEnd) {};
			\draw [draw,-] (Output1) -- node {} (Output1ArcStart);
			\draw [draw,->] (Output1ArcEnd) -- node  [pos=0.5]{\large $ y$} (MonitoringSignal1Below);
			
			\node [input, right of= MonitoringSignal1, node distance=2.5cm] (Eta1) {};
			\node [input, right of= MonitoringSignal1, node distance=2.4cm] (Eta1Arrow) {};
			\draw [draw,->, MyGreen!90] (MonitoringSignal1) -- node  [pos=0.5]{\large \textcolor{black}{$\eta_1$}} (Eta1Arrow);
			
			\node [input, above of= Observer2, right = 1.22cm, node distance=0.25cm] (StateEstimate2line) {};
			\node [input, right of= StateEstimate2line, node distance=1.02cm] (StateEstimate2) {};
			\draw [draw,->, blue!80](StateEstimate2line) -- node  [pos=0.5]{\large \textcolor{black}{$\hat x_2$}} (StateEstimate2);
			\node [input, above of= Observer2, left = 1.23cm, node distance=0.25cm] (Observer2Input) {};
			\node [input, left of= Observer2Input, node distance=1.08cm] (Observer2InputStart) {};
			\node [input, left of= Observer2InputStart, node distance=0.4cm] (Observer2InputArc) {};
			\node [input, left of= Observer2InputArc, node distance=0.5cm] (Observer2InputLeft) {};
			\node [input, above of= Observer2InputLeft, node distance=2cm] (Observer2InputLine) {};
			\draw[black,-] ([yshift=0cm,xshift =-0 cm]Observer2InputArc) arc (180:0:0.2cm);
			\draw [draw,-](Observer2InputArc) -- node {} (Observer2InputLeft);
			\draw [draw,-](Observer2InputLeft) -- node {} (Observer2InputLine);
			
			\draw [draw,->](Observer2InputStart) -- node  [pos=0.5]{\large \textcolor{black}{}} (Observer2Input);
			\node [input, below of= Observer2, right = 1.22cm, node distance=0.25cm] (MS2lineAbove) {};
			\node [input, below of= MS2lineAbove, node distance=0.25cm] (MS2line) {};
			\node [blockG, right of=MS2line, node distance=3.83cm] (MonitoringSignal2) {
				$\begin{array}{c}
					\textbf{Monitoring} \\ \textbf{variable} \ 2
				\end{array}$
			};
			\node [input, above of= MonitoringSignal2, left = 1.49cm, node distance=0.25cm] (MonitoringSignal2Above) {};
			
			\node [input, right of= MS2lineAbove, node distance=0.93cm] (MS2lineAboveArcStart) {};
			\node [input, right of= MS2lineAboveArcStart, node distance=0.4cm] (MS2lineAboveArcEnd) {};
			\draw [draw,-] (MS2lineAbove) -- node {} (MS2lineAboveArcStart);
			\draw [draw,->] (MS2lineAboveArcEnd) -- node  [pos=0.5]{\large $\hat y_2$} (MonitoringSignal2Above);
			
			\node [input, below of= MonitoringSignal2, left = 1.49cm, node distance=0.25cm] (MonitoringSignal2Below) {};
			\node [input, below of= Observer2Line, node distance=0.75cm] (Output2) {};
			
			\node [input, right of= Output2, node distance=4.65cm] (Output2ArcStart) {};
			\node [input, right of= Output2ArcStart, node distance=0.4cm] (Output2ArcEnd) {};
			\draw [draw,-] (Output2) -- node {} (Output2ArcStart);
			\draw [draw,->] (Output2ArcEnd) -- node  [pos=0.5]{\large $ y$} (MonitoringSignal2Below);
			
			\node [input, right of= MonitoringSignal2, node distance=2.4cm] (Eta2) {};
			\draw [draw,->, MyGreen!90] (MonitoringSignal2) -- node  [pos=0.5]{\large \textcolor{black}{$\eta_2$}} (Eta2);
			
			\node [input, above of= ObserverN1, right = 1.22cm, node distance=0.25cm] (StateEstimateN1line) {};
			\node [input, right of= StateEstimateN1line, node distance=1.03cm] (StateEstimateN1) {};
			\draw [draw,->, blue!80](StateEstimateN1line) -- node  [pos=0.5]{\large \textcolor{black}{$\hat x_{N+1}$}} (StateEstimateN1);
			\node [input, above of= ObserverN1, left = 1.23cm, node distance=0.25cm] (ObserverN1Input) {};
			
			\node [input, left of= ObserverN1Input, node distance=1.06cm] (ObserverN1InputStart) {};
			\node [input, left of= ObserverN1InputStart, node distance=0.4cm] (ObserverN1InputArc) {};
			\node [input, left of= ObserverN1InputArc, node distance=2.9cm] (ObserverN1InputLeft) {};
			\draw[black,-] ([yshift=0cm,xshift =-0 cm]ObserverN1InputArc) arc (180:0:0.2cm);
			\draw [draw,-](ObserverN1InputArc) -- node {} (ObserverN1InputLeft);
			\draw [draw,-](inputLine) -- node {} (ObserverN1InputLeft);

			\draw [draw,->](ObserverN1InputStart) -- node  [pos=0.5]{\large \textcolor{black}{}} (ObserverN1Input);
			\node [input, below of= ObserverN1, right = 1.22cm, node distance=0.25cm] (MSN1lineAbove) {};
			\node [input, below of= MSN1lineAbove, node distance=0.25cm] (MSN1line) {};
			\node [blockG, right of=MSN1line, node distance=3.83cm] (MonitoringSignalN1) {
				$\begin{array}{c}
					\textbf{Monitoring} \\ \textbf{variable} \ N+1
				\end{array}$
			};
			\node [input, above of= MonitoringSignalN1, left = 1.49cm, node distance=0.25cm] (MonitoringSignalN1Above) {};
			\draw [draw,->] (MSN1lineAbove) -- node  [pos=0.8]{\large $\hat y_{N+1}$} (MonitoringSignalN1Above);
			\node [input, below of= MonitoringSignalN1, left = 1.49cm, node distance=0.25cm] (MonitoringSignalN1Below) {};
			\node [input, below of= ObserverN1Line, node distance=0.75cm] (OutputN1) {};
			\draw [draw,->] (OutputN1) -- node  [pos=0.9]{\large $y$} (MonitoringSignalN1Below);
			\draw [draw,-] (ObserverN1Line) -- (OutputN1);
			\node [input, above of= StateEstimate1, node distance=0.15cm] (StateEstimate1above) {};
			\node [input, below of= StateEstimateN1, node distance=0.15cm] (StateEstimateN1below) {};
			
			\node [blockMuxBlueBig, below of= StateEstimate1above, node distance=9.0em] (StateEstimateMux) {};
			
			\node [input, right of= MonitoringSignalN1, node distance=2.4cm] (EtaN1) {};
			\draw [draw,->, MyGreen!90] (MonitoringSignalN1) -- node  [pos=0.5]{\large\textcolor{black}{ $\eta_{N+1}$}} (EtaN1);
			\node [input, above of= Eta1, node distance=0.15cm] (Eta1above) {};
			\node [input, below of= EtaN1, node distance=0.15cm] (EtaN1below) {};
			
			\node [blockMuxGreen, below of= Eta1above, node distance=9.0em] (EtaMux) {};
			
			\node at ($(MonitoringSignal2)!.45!(MonitoringSignalN1)$) {\large \textcolor{MyGreen!90}{\vdots}};
			
			\node [input, below of= Eta1, right = 0.12cm, node distance=1cm] (EtaLine) {};
			\node [input, right of= EtaLine, node distance=1.4cm] (Eta) {};
			\draw [draw,-, MyGreen!90] (EtaLine) -- node  [pos=0.6]{\large \textcolor{black}{$\eta$}} (Eta);
			
			\node [blockR, right of=Eta, node distance=2cm] (SigmaBlock) {
				{$	\sigma \in \operatornamewithlimits{\argmin}\limits_{k \in \{1, \dots, N+1\}} \eta_k$}
			};
			\draw [draw,->, MyGreen!90] (Eta) -- (SigmaBlock);
			\node [input, above of= SigmaBlock, right = 1.55cm, node distance=0.7cm] (SigmaBlockName) {};
			\draw [] (SigmaBlockName) -- node  [pos=0.5]{\textbf{Selection Criterion}} (SigmaBlockName);
			
			\node [input, below of= StateEstimate1, right = 0.12cm, node distance=4cm] (StateEstimateLine) {};
			\node [input, right of= StateEstimateLine, node distance=6.3cm] (StateEstimate) {};
			
			\node [input, right of= StateEstimateLine, node distance=4.88cm] (StateEstimateArcStart) {};
			\node [input, right of= StateEstimateArcStart, node distance=0.4cm] (StateEstimateArcEnd) {};
			\draw[blue!80,-] ([yshift=0cm,xshift =-0 cm]StateEstimateArcStart) arc (180:0:0.2cm);
			\draw [draw,-, blue!80] (StateEstimateLine) -- node {} (StateEstimateArcStart);
			\node [input, left of= StateEstimate, node distance=0.1cm] (StateEstimateArrow) {};
			\draw [draw,->, blue!80] (StateEstimateArcEnd) -- node  [pos=0.5]{\large \textcolor{black}{$\hat x$}} (StateEstimateArrow);
			
			\node [input, above of= StateEstimate, node distance=1cm] (StateEstimate1bisLine) {};
			\node [input, above of= StateEstimate, node distance=0.3cm] (StateEstimate2bisLine) {};
			\node [input, below of= StateEstimate, node distance=1cm] (StateEstimateN1bisLine) {};

			\node [blockMuxBlueSmall, below of= StateEstimate1bisLine, node distance=2.85em] (StateEstimateBisMux) {};
			
			\node [input, right of= StateEstimate1bisLine, node distance=1cm] (StateEstimate1bis) {};
			\node [input, right of= StateEstimate2bisLine, node distance=1cm] (StateEstimate2bis) {};
			\node [input, right of= StateEstimateN1bisLine, node distance=1cm] (StateEstimateN1bis) {};
			\node [input, right of= StateEstimate1bisLine, node distance=0.12cm] (StateEstimate1bisArrow) {};
			\node [input, right of= StateEstimate2bisLine, node distance=0.12cm] (StateEstimate2bisArrow) {};
			\node [input, right of= StateEstimateN1bisLine, node distance=0.12cm] (StateEstimateN1bisArrow) {};
			\draw [draw,-, blue!80] (StateEstimate1bisArrow) -- node  [pos=0.5]{\large \textcolor{black}{$\hat x_1$}} (StateEstimate1bis);
			\draw [draw,-, blue!80] (StateEstimate2bisArrow) -- node  [pos=0.5]{\large \textcolor{black}{$\hat x_2$}} (StateEstimate2bis);
			\draw [draw,-, blue!80] (StateEstimateN1bisArrow) -- node  [pos=0.6]{\large \textcolor{black}{$\hat x_{N+1}$}} (StateEstimateN1bis);
			\draw [fill=black] (StateEstimate1bis) circle (2pt);
			\draw [fill=black] (StateEstimate2bis) circle (2pt);
			\draw [fill=black] (StateEstimateN1bis) circle (2pt);
			
			\node at ($(StateEstimate2bis)!.4!(StateEstimateN1bis)$) {\large \vdots};
			
			\node [input, right of= StateEstimate, node distance=2.5cm] (Switching) {};
			\draw [fill=black] (Switching) circle (2pt);
			\draw [draw,-] (StateEstimate1bis) -- (Switching);
			
			\node [input, below of= SigmaBlock, node distance=1.7cm] (SigmaSwitch){};
			\draw [draw,->] (SigmaBlock) -- node  [pos=0.5]{\large $\sigma$} (SigmaSwitch);
			
			\draw[black,-, densely dotted, line width=0.25mm] ([yshift=.5cm,xshift = -0.4cm]Switching) arc (115:245:0.45cm);
			\draw [black,-latex] (17.10,-0.23)to (17.20,-0.17) ;
			\draw [black,-latex] (17.10,-1.08)to (17.20,-1.14) ;
			
			\node [input, right of= SigmaBlock, node distance=2.8cm] (EtaSigma){};
			\draw [draw,->, MyGreen!90] (SigmaBlock) -- node  [pos=0.5]{\large \textcolor{black}{$\eta_\sigma$}} (EtaSigma);
			
			\node [input, right of= Switching, node distance=2.6cm] (hatXSigma){};
			\draw [draw,->, blue!80] (Switching) -- node  [pos=0.5]{\large \textcolor{black}{$	\hat x_\sigma$}} (hatXSigma);
			
			\node [input, left of= EtaSigma, node distance=0.4cm] (EtaSigmaFeedback){};
			\node [input, below of= EtaSigmaFeedback, node distance=4cm] (EtaSigmaFeedbackBelow1){};
			\node [input, below of= EtaSigmaFeedback, node distance=6.5cm] (EtaSigmaFeedbackBelow2){};
			\draw [draw,-, densely dotted, line width=0.2mm, MyGreen!90] (EtaSigmaFeedback) -- (EtaSigmaFeedbackBelow2);
			
			\node [input, below of= MonitoringSignal2, right = 0.3cm, node distance=3cm] (EtaSigmaFeedbackMS2){};
			\draw [draw,-, densely dotted, line width=0.2mm, MyGreen!90] (EtaSigmaFeedbackBelow1) -- (EtaSigmaFeedbackMS2);
			\node [input, right of= MonitoringSignal2, below = 0.63cm, node distance=0.3cm] (MonitoringSignal2right){};
			\draw [draw,->, densely dotted, line width=0.2mm, MyGreen!90] (EtaSigmaFeedbackMS2) -- node  [pos=0.9]{\large \textcolor{black}{$\eta_\sigma$}} (MonitoringSignal2right);
			
			\node [input, below of= MonitoringSignalN1, right = 0.3cm, node distance=1.5cm] (EtaSigmaFeedbackMSN1){};
			\draw [draw,-, densely dotted, line width=0.2mm, MyGreen!90] (EtaSigmaFeedbackBelow2) -- (EtaSigmaFeedbackMSN1);
			\node [input, right of= MonitoringSignalN1, below = 0.63cm, node distance=0.3cm] (MonitoringSignalN1right){};
			\draw [draw,->, densely dotted, line width=0.2mm, MyGreen!90] (EtaSigmaFeedbackMSN1) -- node  [pos=0.3]{\large \textcolor{black}{$\eta_\sigma$}} (MonitoringSignalN1right);
			
			\node [input, left of= hatXSigma, node distance=0.7cm] (hatXSigmaFeedback){};
			\node [input, below of= hatXSigmaFeedback, node distance=1.3cm] (hatXSigmaFeedbackBelow1){};
			\node [input, below of= hatXSigmaFeedback, node distance=3.8cm] (hatXSigmaFeedbackBelow2){};
			\draw [draw,-, densely dotted, line width=0.2mm, blue!80] (hatXSigmaFeedback) -- (hatXSigmaFeedbackBelow2);
			
			\node [input, below of= Observer2, right = 0.3cm, node distance=3.1cm] (hatXSigmaFeedback2){};
			\draw [draw,-, densely dotted, line width=0.2mm, blue!80] (hatXSigmaFeedbackBelow1) -- (hatXSigmaFeedback2);
			\node [input, right of= Observer2, below = 0.63cm, node distance=0.3cm] (Observer2right){};
			\draw [draw,->, densely dotted, line width=0.2mm, blue!80] (hatXSigmaFeedback2) -- node  [pos=0.8]{\large \textcolor{black}{$\hat x_\sigma$}} (Observer2right);
			
			\node [input, below of= ObserverN1, right = 0.3cm, node distance=1.5cm] (hatXSigmaFeedbackN1){};
			\draw [draw,-, densely dotted, line width=0.2mm, blue!80] (hatXSigmaFeedbackBelow2) -- (hatXSigmaFeedbackN1);
			\node [input, right of= ObserverN1, below = 0.63cm, node distance=0.3cm] (ObserverN1right){};
			\draw [draw,->, densely dotted, line width=0.2mm, blue!80] (hatXSigmaFeedbackN1) -- node  [pos=0.3]{\large \textcolor{black}{$\hat x_\sigma$}} (ObserverN1right);
			
			\draw[black,-] ([yshift=0cm,xshift =-0 cm]MS1lineAboveArcStart) arc (180:0:0.2cm);
			\draw[black,-] ([yshift=0cm,xshift =-0 cm]Output1ArcStart) arc (180:0:0.2cm);
			\draw[black,-] ([yshift=0cm,xshift =-0 cm]MS2lineAboveArcStart) arc (180:0:0.2cm);
			\draw[black,-] ([yshift=0cm,xshift =-0 cm]Output2ArcStart) arc (180:0:0.2cm);
			
		\end{tikzpicture}
	\end{center}
	\caption{Block diagram representing the system architecture with $\eta:= (\eta_1, \dots \eta_{N+1})$, $\hat x:= (\hat x_1, \dots, \hat x_{N+1})$.}
	\label{Fig:blockDiagram}
\end{figure*}
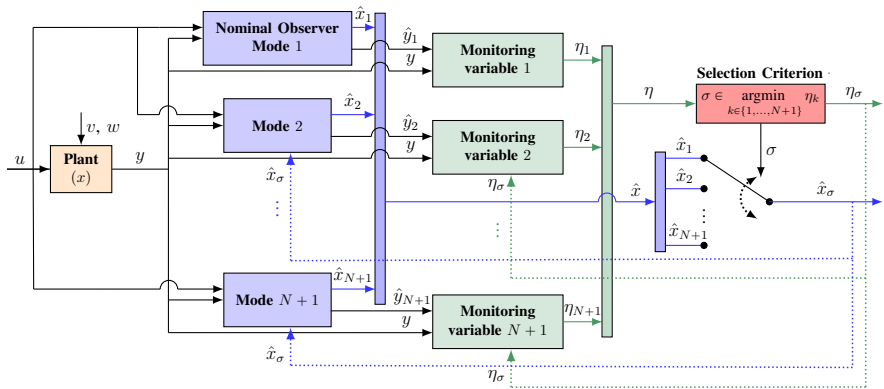

The hybrid estimation scheme we propose consists of the following elements, see Fig.~\ref{Fig:blockDiagram}:
\begin{itemize}[leftmargin=5.5mm]
	\item \textit{nominal observer} given in \eqref{eq:observerNominal}; 
	\item $N$ additional dynamical systems of the form \eqref{eq:observerNominal} but with a different output injection gain, where $N\in\Z_{>0}$. 
	Each of these systems, as well as the nominal observer, is called \textit{mode} for the sake of convenience;
	\item \textit{$N$+1 monitoring variables} used to evaluate the performance of each mode of the multi-observer;
	\item \textit{a selection criterion}, that switches between the state estimates produced by the different modes exploiting the performance knowledge given by the monitoring variables;
	\item \textit{a reset rule}, that defines how the estimation scheme may be updated when the selected mode switches. 
\end{itemize}
All these elements together form the hybrid multi-observer. We describe each component in the sequel.

\subsection{Additional modes} \label{ObserverLikeSystemsSubsection} 
We consider $N$ additional dynamical systems, where the integer $N \in \Zp$ is arbitrarily selected by the user. 
These $N$ extra systems are of the form of \eqref{eq:observerNominal} but with a different output injection gain, i.e., for any $k\in \{2, \dots, N+1\}$, the $k^{\text{th}}$ mode of the multi-observer is given by
\begin{equation}
	\begin{aligned}
		\dot{\hat{x}}_k  &= f_o(\hat{x}_k,u, L_k (y-\hat{y}_k))\\
		\hat{y}_k  &=  h(\hat{x}_k, 0),
	\end{aligned}
	\label{eq:observer}
\end{equation}
where $\hat{x}_k \in \R^{n_x}$ is the $k^{\text{th}}$ mode state estimate, $\hat{y}_k \in \R^{n_y}$ is the $k^{\text{th}}$ mode output and $L_k \in \R^{n_{L_1} \times{n_y}}$ is its gain. 
It is important to emphasize that we make no assumptions on the convergence properties of the solution to system \eqref{eq:observer} contrary to observer \eqref{eq:observerNominal}. There is thus full freedom for selecting the gains $L_k \in \R^{{n_{L_1}}\times n_y}$, with $k \in \{2, \dots, N+1\}$. 
We will elaborate more on the choice of the $L_k$'s in Section~\ref{DesignGuidelinesSection}.
\begin{rem}
	There is also full freedom in the choice of the initial conditions of all modes in \eqref{eq:observerNominal} and \eqref{eq:observer}. 
	We can therefore select all of them equal, but this is not necessary for the stability results in Section~\ref{Main result}  to hold. See Remark~\ref{REM:etaInitCond_Remark} in the sequel for more details. 
\end{rem}
\begin{rem}
The nominal observer \eqref{eq:observerNominal} and the additional modes \eqref{eq:observer} have constant gains $L_k$, but it is also possible to consider time-varying gains $L_k(t)$. 
In this case, if all the gains are uniformly bounded, i.e. there exists $\mathcal{M}>0$ such that $|L_k(t)|\leq \mathcal{M}$ for all $t\geq0$ and all $k\in \{1,\dots,  N+1\}$, then
  the results in this paper hold \emph{mutatis mutandis}. In the numerical example in Section \ref{Example_Battery} one of the modes is an extended Kalman filter, which thus has a time-varying gain.
  We recall that, in general, the extended Kalman filter does not ensure a global convergence property of the estimation error. However, this is not an issue since no stability guarantee is required for the additional modes. 
	\label{Rem:TimeVaryingGains}
\end{rem}

\subsection{Monitoring variables}\label{MonitoringVariables}

Given the $N+1$ modes, our goal is now to find a way to select the ``best" between them, namely the one providing a better estimate, possibly improving the estimation given by the nominal observer~\eqref{eq:observerNominal}.
 Ideally, the criterion used to evaluate the performance of each mode would depend on the estimation errors  $e_k = x- \hat{x}_k$, with $k \in \{1, \dots, N+1\}$. 
However, since the state $x$ is unknown, $e_k$ is unknown and any performance criterion involving $e_k$ would not be implementable. As a consequence, as done in other contexts, see e.g.,  \cite{willems2004deterministic, na2017adaptive}, we rely on the knowledge of the output $y$ and the estimated outputs $\hat{y}_k$ for $k\in \{1,\dots, N+1\}$. 
In particular, inspired by \cite{willems2004deterministic}, in order to evaluate the performance of each mode,  we 
introduce the $N+1$ monitoring variables $\eta_k\in \R_{\geq 0}$ for any $k\in\{1, \dots, N+1\}$, with dynamics given by 
%
%
\begin{equation}
	\begin{aligned}
		\dot{\eta}_k &= -\nu\eta_k + (y-\hat{y}_k)^\top(\Lambda_1 + L_k^\top\Lambda_2L_k)(y-\hat{y}_k)\\ 
		&=: g(\eta_k, L_k, y, \hat{y}_k ), 
	\end{aligned}
	\label{eq:etaDynamics}
\end{equation}
with 
$\Lambda_1 \in \mathbb{S}^{n_y}_{\geq 0}$ and $\Lambda_2 \in \mathbb{S}^{n_x}_{\geq 0}$ with at least one of them positive definite and $\nu \in(0,\alpha]$ a design parameter, where $\alpha$ comes from Assumption~\ref{NominalAssumption}. 
The term $(y-\hat{y}_k)^\top\Lambda_1(y-\hat{y}_k)$ in \eqref{eq:etaDynamics} is related to the output estimation error, while $(y-\hat{y}_k)^\top L_k^\top\Lambda_2L_k(y-\hat{y}_k)$ takes into account the correction effort of the observer, also called latency in \cite{willems2004deterministic}.
Note that the monitoring variable $\eta_k$ in \eqref{eq:etaDynamics} for all $k \in \{1 \dots, N+1\}$ is implementable since we have access to the output $y$ and all the estimated outputs $\hat{y}_k$ at all time instants. 
The monitoring variables $\eta_k$, with $k \in \{1, \dots, N+1\}$, provide evaluations of the performance of all the modes of the multi-observer. 
Indeed, 
by integrating \eqref{eq:etaDynamics} between time $0$ and $t\in \R_{\geq 0}$, we obtain that for any $k \in \{1, \dots, N+1\}$, for any initial condition $\eta_k (0) \in \R_{\geq 0}$, 
for any $y, \hat{y}_k \in \mathcal{L}_{\R^{n_y}}$, and any $t\geq 0$, 
\begin{equation}
	\begin{aligned}
		\eta_k(t) = & \ e^{-\nu t}\eta_k(0) + 
		\int_{0}^{t}e^{-\nu(t-\tau)}\left( (y(\tau)-\hat{y}_k(\tau))^\top(\Lambda_1 \right. \\
		&+  \left. L_k^\top\Lambda_2L_k)(y(\tau)-\hat{y}_k(\tau))   \right)d\tau.
	\end{aligned}
	\label{eq:etaDynamicsTime}
\end{equation}
Equation \eqref{eq:etaDynamicsTime} is a finite-horizon discounted cost, which depends on the output estimation error.


The results we will show in this paper apply for any $\Lambda_1 \in \mathbb{S}^{n_y}_{\geq 0}$ and $\Lambda_2 \in \mathbb{S}^{n_x}_{\geq 0}$ with at least one of them being positive definite. However, their tuning impacts when a switch of the selected mode occurs and which mode is chosen. Indeed,  $\Lambda_1$ and $\Lambda_2$ are the weights of the two terms in \eqref{eq:etaDynamics} and thus their values reflect how much we take into account the output estimation error and the correction effort of the observer in the monitoring variables. 
In particular, selecting $\Lambda_1$ with a norm bigger than $\Lambda_2$ implies that we weight more the term related to the output estimation error compared to the term related to the correction effort of the observer in the design of the monitoring variables and vice-versa.
	In addition, note that, $\Lambda_2$ multiplies the mode gains $L_k$ and thus it implicitly considers also the effect of the measurement noise in the estimation error.

\subsection{Selection criterion}\label{SelectionCriterion}
Based on the monitoring variables $\eta_k$, with $k \in \{1, \dots, N+1\}$, 
we define a criterion to select the state estimate to look at. We use a signal $\sigma:\R_{\geq0} \to \{1, \dots, N+1\}$ for this purpose, and we denote the selected state estimate mode $\hat x_\sigma$ and the associated monitoring variable $\eta_\sigma$. 
The criterion consists in selecting the mode with the minimal monitoring variable, which implies minimizing the cost~\eqref{eq:etaDynamicsTime} over the modes $k \in \{1, \dots, N+1\}$. When several modes produce the same minimum monitoring variable at a given time, 
%
	we select the mode, between the ones with the minimum monitoring variables, with the smaller derivative of $\eta_k$  (which is given by $g(\eta_k, L_k, y, \hat{y}_k)$ from \eqref{eq:etaDynamics}). Moreover, if two or more modes have the same minimum monitoring variable and the same minimum derivative of the monitoring variable, then the proposed technique selects randomly one of them and this is not an issue to obtain the results in 
	Sections~\ref{Main result}, \ref{CompletenessOfSolutionsAndDwellTime} and~\ref{PerformanceImprovement}.
Thus, we switch the selected mode only when there exists $k \in \{1,\dots, N+1\} \setminus \{\sigma\}$ such that\footnote{In \cite{petri2022towards} a switch of the selected mode occurs when there exists $k \in \{1,\dots, N+1\} \setminus \{\sigma\}$ such that $\eta_k \leq \mathfrak{e} \eta_{\sigma}$, where the parameter  $\mathfrak{e} \in  (0,1]$ was introduced to mitigate the occurrence of infinitely fast switching. However, this does not allow to rule out the Zeno phenomenon. To solve this issue, we propose a different jump map in this work, which does not require introducing parameter $\mathfrak{e}$ in the switching criterion as in \cite{petri2022towards}.
} $\eta_k \leq \eta_{\sigma}$. 
%
In that way, at the initial time $t_0 = 0$,  we take $\sigma(0) \in \operatornamewithlimits{\argmin}\limits_{k \in \Pi}
	(g(\eta_k(0), L_k, y(0), \hat{y}_k(0)))$, where $\eta:= \{\eta_1, \dots, \eta_{N+1}\}$ and $\Pi(\eta):= \operatornamewithlimits{\argmin}\limits_{k \in \{ 1, \dots, N +1\} \setminus \{\sigma\}} \eta_k$,  for all $\eta \in \R_{\geq 0}^{N+1}$. Then, $\sigma$ is kept constant, i.e., $\dot \sigma (t) =0$ for all $t \in(0, t_{1})$, with $t_{1}:= \inf \{t \geq 0: \exists k \in \{1,\dots, N+1\} \setminus \{\sigma(t)\}  \text{ such that } \eta_k(t) \leq \eta_{\sigma(t)}(t)\}$. At time $t_1$, we switch the selected mode according to $\displaystyle \sigma(t_{1}^+) \in \operatornamewithlimits{\argmin}\limits_{k \in \Pi}
	(g(\eta_k(t_1), L_k, y(t_1), \hat{y}_k(t_1)))$.
We repeat these steps iteratively and we denote with $t_i \in \R_{\geq 0}$, $i \in \Zp$ the $i^{\textnormal{th}}$ time when the selected mode changes (if it exists), i.e.,  $t_{i}:= \inf \{t \geq t_{i-1}: \exists k \in \{1,\dots, N+1\} \setminus \{\sigma(t)\}  \text{ such that } \eta_k(t) \leq \eta_{\sigma(t)}(t)\}$. Consequently, for all $i\in \Zp$, $\dot \sigma(t) = 0$ for all $t \in (t_{i-1}, t_{i})$ and 
\begin{equation}
\sigma(t_{i}^+) \in \operatornamewithlimits{\argmin}\limits_{k \in \Pi}(
g(\eta_k(t_i), L_k, y(t_i), \hat{y}_k(t_i))),
\label{eq:sigma_update}
\end{equation}
where we recall that $\eta= \{\eta_1, \dots, \eta_{N+1}\}$ and $\Pi(\eta)= \operatornamewithlimits{\argmin}\limits_{k \in \{ 1, \dots, N +1\} \setminus \{\sigma\}} \eta_k$,  for all $\eta \in \R_{\geq 0}^{N+1}$.
We also argue that, implementing \eqref{eq:sigma_update} online, which requires the knowledge of the derivative of the monitoring variables, is not an issue.

\begin{rem}
	The scheme proposed in this paper works for any initial condition  $\eta_k(0) \in \R_{\geq 0}$, 
	for all $k \in \{1, \dots, N+1\}$, which corresponds to the initial cost of each mode of the multi-observer. 
	Consequently, the choice of $\eta_k(0)$ is an extra degree of freedom that can be used to initially penalize the modes when there is a prior knowledge of which mode should be initially selected, as done in \cite{petri2023state}. 
	Conversely, in the case where there is no prior knowledge on which mode should be chosen at the beginning, all $\eta_k$, with $k \in \{1, \dots, N+1\}$, can be initialized at the same value such that the term $e^{-\nu t}\eta_k(0)$ in \eqref{eq:etaDynamicsTime} is irrelevant for the minimization.
	\label{REM:etaInitCond_Remark}
\end{rem}

\begin{rem}
	The results in Sections~\ref{Main result} and \ref{CompletenessOfSolutionsAndDwellTime} also apply with  $\sigma(t_i^+) \in \operatornamewithlimits{\argmin}\limits_{k \in \{ 1, \dots, N +1\} \setminus \{\sigma\}} \eta_k$ instead of \eqref{eq:sigma_update}, as in \cite{petri2022towards}. To select the mode with the minimum derivative of the monitoring variable, among those with the minimum $\eta_k$, allows us to prove a strict performance improvement in Section~\ref{PerformanceImprovement}.
\end{rem}
%
\subsection{Reset rule}\label{ResetRule}
When a switching occurs, i.e., when a different mode is selected, 
we propose two different options to update the hybrid estimation scheme. The first one, called \textit{without resets}, consists in only updating $\sigma$, and consequently, we only switch the state estimate we are looking at. Conversely, the second option, called \textit{with resets}, consists in not only switching the mode that is considered, but also resetting the state estimates and the monitoring variables of all the modes $k \in \{2, \dots, N+1\}$ to the updated $\hat x_\sigma$ and $\eta_\sigma$, respectively. 
The state estimate and the monitoring variable of the nominal observer~\eqref{eq:observerNominal}, corresponding to mode $1$, are never reset. 
Since the reset option re-initializes the modes states at each switching time, it can be useful when one or more modes are local observers, and thus they guarantee a convergence property of the estimation error only if they are initialized sufficiently close to the origin, or when the null gain is selected for one of the additional modes. 
 However, the reset case requires communication between the modes, which may not be always implementable in practice.

To avoid infinitely fast switching, we introduce a regularization parameter $\varepsilon \in \R_{> 0}$. In particular, when a switch of the selected mode occurs, the value of monitoring variables $\eta_k$, with $k \in \{2, \dots, N+1\}\setminus \{\sigma\}$, is increased by $\varepsilon$, both in the case without and with resets. The idea is to penalize the unselected modes and to allow the selected one to run for some amount of time before a new switch occurs. 
The analysis of the properties of the hybrid time domains of the overall solutions and the existence of a uniform semiglobal average dwell-time are presented Section~\ref{CompletenessOfSolutionsAndDwellTime}. 

We use the parameter $r \in \{0,1\}$ to determine which option is selected, where $r = 0$ corresponds to the case without resets, while $r = 1$ corresponds to the case where the resets are implemented. Note that, the parameter $r$ needs to be chosen off-line and it is fixed for each considered observation process.
When a switch of the considered mode occurs, the state estimate $\hat x_k$ of the $k^{\text{th}}$ mode 
is defined as, at a switching time $t_i \in \R_{\geq 0}$, 
\begin{equation}
	\hat{x}_1(t_i^+) := \hat{x}_1(t_i)
	\label{eq:StateEstimate_plus_nom}
\end{equation}
and, for all $k \in \{2, \dots, N+1\}$,
\begin{equation}
	\begin{aligned}
		\hat x_k(t_i^+) &\in \{(1-r)\hat x_k(t_i) + r \hat x_{k^\star}(t_i):k^\star \in  \operatornamewithlimits{\argmin}\limits_{j \in \Pi} (g(\eta_j(t_i), \\
		& \quad L_j, y(t_i), \hat{y}_j(t_i))) \}\\
		&=:\hat \ell_k (\hat x(t_i), \eta(t_i), L, y(t_i), \hat{y}(t_i)), 
	\end{aligned}
	\label{eq:StateEstimatek_plus}
\end{equation}
where $\hat x:= (\hat x_1, \dots, \hat x_{N+1})$, $\eta= (\eta_1, \dots, \eta_{N+1})$, $L:= (L_1, \dots, L_{N+1})$ and $\hat y:= (\hat y_1, \dots \hat y_{N+1})$. 
%
Similarly, at a switching time $t_i \in \R_{\geq 0}$, the monitoring variables are defined as, 
\begin{equation}
	\eta_1(t_i^+) := \eta_1(t_i),  
	\label{eq:Eta_plus_nom}
\end{equation}
\begin{equation}
	\begin{aligned}
		\eta_\sigma(t_i^+) := \eta_\sigma(t_i)
	\end{aligned}
	\label{eq:Eta_plus_sigma}
\end{equation}
and, for all $k \in \{2, \dots, N+1\} \setminus \{\sigma\}$, 
\begin{equation}
	\begin{aligned}
		&\eta_k(t_i^+) = (1-r)\eta_k + r \eta_{k^\star} + \varepsilon 
	\end{aligned}
	\label{eq:Eta_plus}
\end{equation}
where $\varepsilon \in \R_{> 0}$ and 
$\eta_{k^\star} = \operatornamewithlimits{\min}\limits_{j \in \{ 1, \dots, N +1\} \setminus \{\sigma\}} \eta_j$.  
\color{black}
Note that, if the monitoring variables of more than one mode have the same value and it is the minimum between all the $\eta_k$, with $k \in \{1, \dots, N+1\}$, then, from \eqref{eq:StateEstimatek_plus}, the modes may be reset with different state estimates. 
Merging \eqref{eq:Eta_plus_sigma} and \eqref{eq:Eta_plus} and using the Kronecker delta definition, 
we obtain, at a switching time $t_i \in \R_{\geq 0}$, for all $k \in \{2, \dots, N+1\}$, 
\begin{equation}
	\begin{aligned}
		\eta_k(t_i^+) &= \delta_{k,\sigma} \eta_k + (1-\delta_{k,\sigma})((1-r)\eta_k + r \eta_{k^\star} + \varepsilon)\\
		&=:p_k (\eta(t_i)), 
	\end{aligned}
	\label{eq:Eta_plus_compact}
\end{equation}
where $\varepsilon \in \R_{> 0}$ and
$\eta_{k^\star} = \operatornamewithlimits{\min}\limits_{j \in \{ 1, \dots, N +1\} \setminus \{\sigma\}} \eta_j$.
%
%

We can already note that, with the proposed technique, $\eta_{\sigma(t)}(t) \leq \eta_1(t)$ for all $t \geq 0$, both in the case without and with resets. Therefore, the estimation performance of the proposed hybrid multi-observer is always not worse than the performance of the nominal one according to the monitoring variables that we consider. 
We will study the performance of the estimation scheme in more depth in Section \ref{PerformanceImprovement}.

\subsection{Hybrid model}\label{HybridModel}
To proceed with the analysis of the hybrid estimation scheme presented so far, we model the overall system as a hybrid system of the form of \cite{goebel2012hybrid}, where a jump corresponds to a switch of the selected mode and a possible reset as explained in Section~\ref{ResetRule}. We define the overall state as $q:= (x, \hat{x}_1, \dots, \hat{x}_{N+1},\eta_1, \dots, \eta_{N+1}, \sigma) \in \mathcal{Q}:= \R^{n_x} \times \R^{(N+1)n_x} \times \R_{\geq 0}^{N+1}  \times \{1, \dots, N +1\}$,
and we obtain the hybrid system 
\begin{equation}
	\left\lbrace \
	\begin{aligned}
		\dot{q} &= F(q, u, v, w), \ \ \ \ \ &&q \in \mathcal{C}\\
		q^{+} &\in G(q), \ \ \ \ \ &&q \in \mathcal{D},
	\end{aligned}
	\right.
	\label{eq:HybridSystemGeneral}
\end{equation}
where flow map is defined as, for any $q \in \mathcal{C}$, $u \in \mathcal{U}$, $v \in \mathcal{V}$ and $w \in \mathcal{W}$, from \eqref{eq:system}, \eqref{eq:observerNominal}, \eqref{eq:observer}, \eqref{eq:etaDynamics},
	$$
	\begin{aligned}
		F &:= (f_p, f_{o,1}, \ldots, f_{o,N+1}, g_1, \ldots, g_{N+1}, 0)\\
		G &:= (x, \hat x_1, \hat\ell_2 \ldots, \hat\ell_{N+1}, \eta_1, p_2, \ldots, p_{N+1}, \operatornamewithlimits{\argmin}\limits_{k \in \Pi} g_k )
	\end{aligned}
	$$
	with the short notation $f_{o,k} = f_o(\hat x_k, u, L_k(y-y_k))$, $g_k = g(\eta_k, L_k, y, y_k)$, $\hat\ell_k = \hat\ell_k (\hat x, \eta, L_k, y, \hat{y}_k)$ and $p_k = p_k (\eta)$, for all $k \in \{1, \dots, N+1\}$,  where $\Pi(q)= \operatornamewithlimits{\argmin}\limits_{k \in \{ 1, \dots, N +1\} \setminus \{\sigma\}} \eta_k$ for all $q \in \mathcal{D}$. 
In view of Section \ref{SelectionCriterion}, the flow and jump sets $\mathcal{C}$ and $\mathcal{D}$ in \eqref{eq:HybridSystemGeneral} are defined as 
\begin{align}
	\mathcal{C} := \left\{q \in \mathcal{Q}: \forall k \in \{1, \dots, N+1\} \:  \ \eta_k \geq  \eta_\sigma \right\}\!,	\label{eq:flowSet} \\
	\mathcal{D} := \left\{q \in \mathcal{Q}: \exists k \in \{1, \dots, N+1\} \setminus \{\sigma\} \: \ \eta_k \leq \eta_\sigma \right\}\!.
	\label{eq:jumpSet}
\end{align}
\begin{rem}
	The state estimate $\hat x_\sigma$ of the hybrid multi-observer is subject to jumps, which may not be suitable in some applications. For this reason, as done in \cite{petri2023state} for the state estimation of Li-Ion batteries, it is possible to add a filtered version of $\hat x_\sigma$, denoted $\hat x_f$, whose dynamics between two successive switching instants is $\dot{\hat{x}}_\sigma = -\zeta \hat x_f + \zeta \hat x_\sigma$, where $\zeta >0$ is an additional design parameter and,  $\hat x_f$ does not change at switching times $t_i \in \R_{\geq 0}$, i.e., $\hat x_f(t_i^+) =  \hat x_f(t_i)$. Note that, when the filtered version of the hybrid multi-observer state estimate is also considered, similar stability results can be proved, see \cite[Section 5.3.6]{petri2023thesis}. 
\end{rem}

\color{blue}
\section{Design guidelines}\label{DesignGuidelinesSection}
\color{black}
We summarize the procedure to follow to design the hybrid estimation scheme. 
\begin{enumerate}
	\item  Design the nominal observer \eqref{eq:observerNominal} such that Assumption~\ref{NominalAssumption} holds. 
	\item  Select $N$ gains $L_2, \dots, L_{N+1}$ 
	for the $N$ additional modes in  \eqref{eq:observer}.
	\item  Implement in parallel the $N+1$ modes of the multi-observer.
	\item Generate the monitoring variables $\eta_k$, with $k \in \{1, \dots, N+1\}$. 
	\item Evaluate the signal $\sigma$.
	\item  Select $\varepsilon \in \R_{> 0}$ and run the hybrid scheme without or with resets.
	\item  $\hat x_\sigma$ is the state estimate to be considered for estimation purpose. 
	\label{DesignStepsAlgorithm}
\end{enumerate}

There is a lot of flexibility in the number of additional modes $N$ and the selection of the gains $L_k$, with $k \in \{2, \dots, N+1\}$. This allows to address the different trade-offs of the state estimation of nonlinear systems.
To present a systematic procedure to design the additional gains is very challenging in view of the generality of the considered classes of systems and observers. However, we can provide general principles to be followed.
\subsection{General guidelines for gains selection}\label{GuidelinesNullGainAndOptimization}
\subsubsection{Null gain}
We can always select one of the additional mode gains as the null gain, namely $L_k = 0_{n_x \times n_y}$, for some $k \in \{2, \dots N+1\}$. This choice will produce an open-loop observer, which therefore typically does not have any stability guarantee. However, it is the best gain choice to annihilate the effect of the measurement noise on the estimation error and thus this gain can be very useful to improve the estimation performance, especially when the resets are implemented, as we will show on numerical examples in Section~\ref{Example}. 

\subsubsection{Optimization-based design}\label{Optimization}
The additional modes gains can be designed by solving optimization problems off-line for different possible sets of initial conditions and classes of inputs and disturbances. Typically, the gains obtained in this way do not guarantee any stability property of the estimation error. However, this is not required for the proposed hybrid estimation scheme, as testified by Theorem \ref{Prop:LyapunovSolutionProposition}. Among the different cost functions that can be considered to solve the optimization problem, one possible option is described in the following. Given $x_0, \hat x_0 \in \R^{n_x}$, $u\in \mathcal{U}$,  $v\in \mathcal{V}$ and $w\in \mathcal{W}$ we define, 
for a given $t \in \R_{\geq 0}$, 
\begin{equation}
	\begin{aligned}
	\mathcal{J}(x_0, \hat x_0, u, v, w, L_k, t) &:=\\
	 \int_{0}^{t} e^{-\vartheta s} (x(s)& - \hat{x}_k(s))^\top Q (x(s) - \hat{x}_k(s)) ds, 
	\label{eq:optimizationCost}
	\end{aligned}
\end{equation}
where $\vartheta \in \R_{\geq 0}$, $Q \in \mathbb{S}^{n_x}_{\geq 0}$ is a weight matrix, $k \in \{2, \dots, N+1\}$, and $x(t)$, $\hat x_k(t)$ represent the solutions to systems \eqref{eq:system} and \eqref{eq:observerNominal}, with $x(0) = x_0$ and $\hat x_k(0) = \hat x_0$, for all $t\in \R_{\geq 0}$, respectively. 
The optimization problem to solve to obtain the gain $L_k \in \R^{n_x \times n_y}$ is given by
\begin{equation}
	\min\limits_{L_k \in \R^{n_x \times n_y}} \max\limits_{v \in \mathcal{V}, w \in \mathcal{W}} \mathcal{J}(x_0, \hat x_0, u, v, w, L_k, t). 
	\label{eq:optimizationProblem}
\end{equation}
Cost \eqref{eq:optimizationCost} is a quadratic cost of the state estimation error, which is available off-line in simulations. By solving the optimization problem \eqref{eq:optimizationProblem}, we obtain the gain $L_k$, $k \in \{2,\dots, N+1\}$, that minimizes cost \eqref{eq:optimizationCost} for the worst case scenario for the disturbance $v \in \mathcal{V}$ and measurement noise $w \in \mathcal{W}$, for the considered input $u \in \mathcal{U}$ and initial conditions $x_0$ and $\hat x_0 \in \R^{n_x}$ for systems \eqref{eq:system} and \eqref{eq:observerNominal}, respectively.  
By solving \eqref{eq:optimizationProblem} for various initial conditions, inputs $u$ and classes of disturbances and noises and final time $t$, we obtain a bank of observer gains, which can be used in each mode of the multi-observer.

\subsubsection{Adjusting $L_1$}
	 We can also select the additional gains in a neighborhood of the nominal one, or to scale the nominal gain by some factors. 
	This gain selection will produce systems with different behaviors and switching between them should allow an improvement of the estimation performance.

\subsection{Exploiting system and nominal observer structure and/or behaviour}\label{GainTuningSpecificCase}
%
The gain selection can be also done by exploiting the knowledge of the system and observer structures.
For example, when the nominal observer \eqref{eq:observerNominal} is a high-gain observer,  see e.g., \cite{khalil2014high, astolfi2015high} or, more generally, an infinite gain margin observer \cite[Section 3.4]{bernard2022observer}, we typically need to select a very large gain based on a conservative bound to ensure Assumption~\ref{NominalAssumption}, which would result in 
fast convergence of the estimation error, but, unfortunately, it will be very sensitive to measurement noise. In this case, to overcome the conservatism of the theory, an option is to select the $L_k$ gains (much) smaller than the nominal one, even though there is no convergence proof for these choices, in order to obtain a state estimate which is more robust to measurement noise. This is the approach followed in Section~\ref{Example_VanDerPol} on an example. 

Another possible approach is to select the additional gains $L_k$'s considering the behavior of the nominal observer in simulation and choose them based on the properties we want to improve. For instance, similarly to the case where the nominal observer is an high-gain observer, when the convergence speed of the nominal observer is satisfactory, but its estimation error is very sensitive to noises, the gains $L_k$'s should be selected smaller than the nominal one $L_1$. On the other hand, if the convergence speed of the estimation error of the nominal observer is too slow, the additional gains may be chosen bigger than $L_1$. This approach to select the additional gains was used in \cite{petri2023state}, where the hybrid multi-observer presented in \cite{petri2022towards} was implemented to improve the estimation performance of a electrochemical Li-Ion battery model.  

\section{Stability guarantees}\label{Main result}
The goal of this section is to prove that the proposed hybrid estimation scheme satisfies an input-to-state stability property. 
Even though the nominal observer satisfies an input-to-state stability property by Assumption \ref{NominalAssumption}, it is not obvious that so does system \eqref{eq:HybridSystemGeneral}-\eqref{eq:jumpSet}, as the extra modes are designed with no convergence guarantees.

\subsection{Input-to-state stability}\label{StabilitySubsectionSolutions}
In the next theorem we prove that system \eqref{eq:HybridSystemGeneral}-\eqref{eq:jumpSet} satisfies an input-to-state stability property. 
\begin{thm}
	Consider system \eqref{eq:HybridSystemGeneral}-\eqref{eq:jumpSet} and suppose Assumptions~\ref{NominalAssumption}-\ref{ASS:ass1} hold. 
	Then there exist $\beta_U \in \KL$ and $\gamma_U \in \Kinf$ such that for any input $u \in \mathcal{L_U}$, disturbance input $v \in \mathcal{L_V}$ and measurement noise $w \in \mathcal{L_W}$, any solution $q$ satisfies 
	\begin{equation}
		\begin{array}{l}
			|(e_1(t,j), \eta_1(t,j), e_{\sigma} (t,j), \eta_{\sigma}(t,j))|  \\[0.5em]
			\qquad \leq \beta_U(|(e(0,0), \eta(0,0))|,t) + \gamma_U(\norm{v}_{[0,t]} + \norm{w}_{[0,t]})
		\end{array}
		\label{eq:LyapunovSolutionTheoremEquation}
	\end{equation}
	for all $(t,j) \in \dom q$, with $e:= (e_1, \dots, e_{N+1})$ and $\eta:= (\eta_1, \dots, \eta_{N+1})$.
	\label{Prop:LyapunovSolutionProposition}
\end{thm}
%
Theorem~\ref{Prop:LyapunovSolutionProposition} guarantees a two-measure input-to-state stability property \cite{cai2007smooth}. 
In particular, \eqref{eq:LyapunovSolutionTheoremEquation} ensures that $e_1$, $\eta_1$,  $e_\sigma$ and $\eta_\sigma$ converge to a neighborhood of the origin, whose ``size'' depends on the $\mathcal{L}_\infty$ norm of $v$ and $w$. Note that we do not guarantee any stability property for the modes $k \neq \sigma$, but this is not needed for the convergence of the hybrid observer estimation error $e_\sigma$. 
Hence, the convergence of the estimated state vector of the selected mode is guaranteed by Theorem~\ref{Prop:LyapunovSolutionProposition}. 
We would like to emphasize 
	that \eqref{eq:LyapunovSolutionTheoremEquation} does not inform us about the performance improvement of the hybrid scheme, this will be addressed in Section \ref{PerformanceImprovement}.

The proof of Theorem~\ref{Prop:LyapunovSolutionProposition} is omitted for
	space reasons and can be found in \cite[Proof of Theorem 5.1]{petri2023thesis}.

\subsection{Lyapunov properties}\label{LyapunovSubsection}
In this section we state the Lyapunov properties,  which are employed to prove Theorem~\ref{Prop:LyapunovSolutionProposition}. 
%
Based on  Assumption~\ref{NominalAssumption}, we first prove an input/output-to-state stability property \cite{sontag2008input} for the generic estimation error system $e := x- \hat{x} \in \R^{n_x}$ associated with \eqref{eq:system} and \eqref{eq:observer}, whose dynamics is defined as 
	$	\dot e = f_p(x,u,v)- f_o(\hat{x},u,  L(y - \hat{y}))$ 
with $L\in \R^{{n_{L_1}} \times n_y}$.
\begin{lem}\label{PropositionOSSobservers}
	Suppose Assumption~\ref{NominalAssumption} holds. 
	Then, for any $x \in \R^{n_x}$, $u \in  \mathcal{U}$, $v \in  \mathcal{V}$, $w \in  \mathcal{W}$, $\hat{x} \in \R^{n_x}$ and any $L \in \R^{{n_{L_1}} \times n_y}$, 
	\begin{equation}
		\begin{array}{l}
			\left\langle \nabla V(e),f_p(x,u,v)- f_o(\hat{x},u,  L(y - \hat{y})) \right\rangle \\
			\quad \leq -\alpha V(e) + \psi_1(|v|) + \psi_2(|w|) + \gamma\norm{L-L_1}^2 |y- \hat{y}|^2,\\
		\end{array}
		\label{eq:PropositionOSS}
	\end{equation}
	with 
	$\hat{y} = h(\hat{x}, 0) \in \R^{n_{y}}$ and $\alpha, \psi_1, \psi_2, \gamma, V$ 
	come from Assumption~\ref{NominalAssumption}. 
	\label{Lemma:IOSSlemma}
\end{lem}
\noindent\textbf{Proof:} 
Let $x \in \R^{n_x}$, $u \in  \mathcal{U}$, $v \in  \mathcal{V}$, $w \in  \mathcal{W}$, $\hat{x} \in \R^{n_x}$ and $L \in \R^{{n_{L_1}} \times n_y}$, we have that
$\left\langle \nabla V(e), f_p(x,u,v)- f_o\big(\hat{x},u, L(y - \hat{y})\big) \right\rangle  
= \left\langle \nabla V(e), f_p(x,u,v) \right.
 \left. - f_o\big(\hat{x},u, L_1(y - \hat{y}) - L_1(y - \hat{y})+ L \right.$  $\left. (y - \hat{y})\big) \right\rangle 
 =\left\langle \nabla V(e), f_p(x,u,v) \right.  \left. - f_o\big(\hat{x},u, L_{1}(y - \hat y)  \right.$  $\left. +(L -L_1)(y - \hat{y})\big) \right\rangle.$
Applying Assumption~\ref{NominalAssumption} with $d = (L-L_1)(y-\hat y)$ we obtain
$\left\langle \nabla V(e), f_p(x,u,v)- f_o\big(\hat{x},u, L(y - \hat{y})\big) \right\rangle  
\leq  -\alpha V(e) +\psi_1(|v|)  + \psi_2(|w|) + \gamma|(L - L_1)(y- \hat{y})|^2$, 
which implies \eqref{eq:PropositionOSS}. We have obtained the desired result. 
\hfill $\blacksquare$
\color{black}

Lemma~\ref{PropositionOSSobservers} implies that, for $e_k:=x-\hat x_k$ for any $k\in \{2, \dots, N+1\}$, the $e_k$-system, which follows from \eqref{eq:system} and \eqref{eq:observer}, satisfies an input/output-to-state property \cite{sontag2008input} with the same Lyapunov function as in Assumption~\ref{NominalAssumption} for any choice for the observer gain $L_k \in \R^{{n_{L_1}} \times n_y}$. 
The major difference between \eqref{eq:NominalAssumptionDerivative} and \eqref{eq:PropositionOSS} is the term $ \gamma||(L - L_1)||^2|y- \hat{y}|^2$ in \eqref{eq:PropositionOSS}, which may have a destabilizing effect and may thus prevent the $e_k$-system to exhibit input-to-state stability properties similar to~\eqref{eq:ISSnominalConverging}. 
In the next proposition, we state Lyapunov properties for system \eqref{eq:HybridSystemGeneral}-\eqref{eq:jumpSet}, whose proof is postponed to Appendix \ref{Proof:Prop1}. 

\begin{prop}
	Suppose Assumptions~\ref{NominalAssumption}-\ref{ASS:ass1} hold.  
	Given any sets of gains  $L_k \in \R^{{n_{L_1}} \times{n_y}}$, with $k \in \{2, \dots, N+1\}$, any $\nu \in (0,\alpha]$, any $\varepsilon > 0$ and any $\Lambda_1 \in \mathbb{S}^{n_y}_{\geq 0}$, $\Lambda_2 \in \mathbb{S}^{n_x}_{\geq 0}$ with at least one of them positive definite, 
	there exist $U:\mathcal{Q} \to \R_{\geq 0}$ locally Lipschitz, and 
	$\underline{\alpha}_U, \overline{\alpha}_U \in \Kinf$, $\alpha_0 \in \R_{> 0}$, $\phi_1, \phi_2 \in \Kinf$, such that
	the following properties hold. 
	\begin{enumerate}[label=(\roman*),leftmargin=.6cm]
		\item  
		$\underline{\alpha}_U(|(e_1, \eta_1, e_\sigma, \eta_\sigma)|) \leq U(q) \leq \overline{\alpha}_U(|( e, \eta)|)$ for any $q \in~\mathcal{Q}$,
		with $e= (e_1, \dots, e_{N+1})$ and $\eta= (\eta_1, \dots, \eta_{N+1})$. 
		\item 
		$U^\circ(q;F(q,u,v,w)) \leq -\alpha_0 U(q) + \phi_1(|v|) + \phi_2(|w|)$ for any $q \in \mathcal{C}$, $u \in \mathcal{U}$, $v \in \mathcal{V}$ and $w \in \mathcal{W}$, such that $F(q,u,v,w) \in T_\mathcal{C}(q)$, where we recall that $U^\circ$ denotes the Clarke generalized directional derivative of $U$, as defined in Section \ref{Introduction}.
		\item 
		$U(\mathfrak{g} ) \leq U(q)$ for any $q \in \mathcal{D}$ and any $\mathfrak{g} \in G(q)$. 
	\end{enumerate}
	\label{THM:LyapunovTheorem}
\end{prop}

Proposition~\ref{THM:LyapunovTheorem} shows the existence of a Lyapunov function $U$ for system \eqref{eq:HybridSystemGeneral}-\eqref{eq:jumpSet}, which is used to prove the input-to-state stability property in Theorem~\ref{Prop:LyapunovSolutionProposition}. 

\section{Properties of the solution domains}\label{CompletenessOfSolutionsAndDwellTime}
An input-to-state stability property is established in Theorem~\ref{Prop:LyapunovSolutionProposition} but nothing is said about the completeness and more generally about the properties of the solutions time domains. In this section, we address these points. 
In Section~\ref{CompletenessOfSolutions}, we show that maximal solutions are complete, while in Section~\ref{DwellTime} we ensure the existence of a uniform semiglobal average dwell-time thereby ruling out Zeno phenomenon. 

\subsection{Completeness of maximal solutions}\label{CompletenessOfSolutions}
The goal of this section is to show that maximal solutions to system \eqref{eq:HybridSystemGeneral}-\eqref{eq:jumpSet} are complete, which means that their domains are unbounded. For this purpose, we need that the system plant \eqref{eq:system} is complete, as stated in the next assumption.  
\begin{ass}
	Any maximal solution to \eqref{eq:system} with $u$ in $\mathcal{L_U}$, $v$ in $\mathcal{L_V}$ and $w$ in $\mathcal{L_W}$ is complete.  
	\label{ASS:forwardCompletenessAssumption}
\end{ass}
Before proving the main result of this section, we show in the next lemma that maximal solutions to the additional modes \eqref{eq:observer} are complete. 

\begin{lem}
	Consider systems~\eqref{eq:system} and~\eqref{eq:observer}. Suppose Assumptions~\ref{NominalAssumption},~\ref{ASS:ass1} and~\ref{ASS:forwardCompletenessAssumption} hold. Then, for any 
	inputs $u \in \mathcal{L_U}$, $v \in \mathcal{L_V}$,  $w \in \mathcal{L_W}$ and $y \in \mathcal{L}_{\R^{n_y}}$, any corresponding maximal solution to \eqref{eq:observer} is complete. 
	\label{Lemma:CompleteAdditionalModes}
\end{lem}

\noindent\textbf{Proof:}
Let $k \in \{2, \dots N+1\}$ and let $x \in \R^{n_x}$, $u \in  \mathcal{U}$, $v \in  \mathcal{V}$, $w \in  \mathcal{W}$, $\hat{x}_k \in \R^{n_x}$ and any $L_k \in \R^{{n_{L_1}} \times n_y}$. From Lemma \ref{Lemma:IOSSlemma}, we have, for all $k \in \{2, \dots, N+1\}$, 
\begin{equation}
	\begin{array}{l}
		\left\langle \nabla V(e_k), f_p(x,u,v)- f_o\big(\hat{x}_k,u, L_k(y - \hat{y}_k)\big) \right\rangle \\
		\leq -\alpha V(e_k) + \psi_1(|v|) + \psi_2(|w|) + \gamma\norm{L_k-L_1}^2 |y- \hat{y}_k|^2\\
		\leq -\alpha V(e_k) + \psi_1(|v|) + \psi_2(|w|) + \theta |y- \hat{y}_k|^2,\\
	\end{array}
	\label{eq:OSSInLemmaCompleteness}
\end{equation}
with  $\theta:= \gamma \max\limits_{k \in \{1, \dots, N+1\}}{\norm{L_k - L_1}^2} \in \R_{\geq 0}$.
Using Assumption~\ref{ASS:ass1} we have $ |y -\hat y_k|^2 = |h(x,w) -h(\hat x_k,0)|^2 \leq \delta_1 V(e_k) +\delta_2|w|^2$, for all $k \in \{2, \dots, N+1\}$. Thus, from \eqref{eq:OSSInLemmaCompleteness} we obtain, 
\begin{equation}
	\begin{array}{l}
		\left\langle \nabla V(e_k),f_p(x,u,v)- f_o\big(\hat{x}_k,u, L_k(y - \hat{y}_k)\big) \right\rangle \\
		\leq-\alpha V(e_k) + \psi_1(|v|) + \psi_2(|w|) + \theta\delta_1V(e_k) + \theta\delta_2|w|^2\\
		= \mathfrak{a} V(e_k) + \psi_1(|v|) + \psi_2^\star(|w|),\\
	\end{array}
	\label{eq:OSSInLemmaCompleteness2}
\end{equation}
with $\mathfrak{a}:= \theta\delta_1- \alpha \in \R$ and $\psi_2^\star: s \mapsto  \psi_2(|s|)+\theta\delta_2|s|^2 \in \Kinf$. 

Let  $u \in \mathcal{L_U}$,  $v \in \mathcal{L_V}$,  $w \in \mathcal{L_W}$  and $x$ and $\hat{x}_k$ be solutions to systems~\eqref{eq:system} and~\eqref{eq:observer} respectively, for  $k \in \{2, \dots, N+1\}$. We have, by definition, $e_k(t)= x(t) - \hat{x}_k(t)$, for all $k \in \{2, \dots, N+1\}$ and all $t \in \dom(x, \hat x_k)$. 
Pick any $k \in \{2, \dots, N+1\}$, 
for all $t \in [0, \infty)$, we have from \eqref{eq:OSSInLemmaCompleteness2},
\begin{equation}
	\frac{d}{dt}V(e_k(t))\leq \mathfrak{a} V(e_k(t)) + \psi_1(|v(t)|) + \psi_2^\star(|w(t)|).
\end{equation}
Applying the comparison principle \cite[Lemma 3.4]{khalil2002nonlinear}, we obtain, for all $t \in [0,\infty)$, 
$	V(e_k(t))
\displaystyle \leq e^{\mathfrak{a} t}V(e_k(0)) 
+ \int_{0}^{t}e^{\mathfrak{a} (t-s)}(\psi_1(v(|s|)) + \psi_2^\star(|w(s)|))ds.$
From \eqref{eq:NominalAssumptionSandwichBound} and the last inequality, $e_k$ cannot blow up in finite time as $V$ is positive definite and the right hand side is finite for any $t \geq 0$. 
Moreover, from Assumption~\ref{ASS:forwardCompletenessAssumption}, $x$ cannot blow up in finite time. Consequently, since $\hat{x}_k = x-e_k$ and both $x$ and $e_k$ cannot explode in finite time, $\hat{x}_k$ cannot as well. 
Thus, for any $k \in \{2, \dots, N+1\}$, any maximal solution to system~\eqref{eq:observer} is complete.
\hfill $\blacksquare$

We are now ready to prove the completeness of maximal solution of system  \eqref{eq:HybridSystemGeneral}-\eqref{eq:jumpSet}.
\begin{prop}
	Under Assumptions \ref{NominalAssumption}, \ref{ASS:ass1} and \ref{ASS:forwardCompletenessAssumption}, for any inputs $u \in \mathcal{L_U}$, $v \in \mathcal{L_V}$,  $w \in \mathcal{L_W}$,  any maximal solution to system \eqref{eq:HybridSystemGeneral}-\eqref{eq:jumpSet} 
	is complete. 
	\label{THM:CompletenessOfMaximalSolutionsTheorem}
\end{prop}
%
\noindent\textbf{Proof:}
We use \cite[Proposition 6]{heemels2021hybrid} to prove Proposition~\ref{THM:CompletenessOfMaximalSolutionsTheorem}. Let $u \in \mathcal{L_U}$, $v \in \mathcal{L_V}$, $w \in \mathcal{L_W}$ and $q$ be a maximal solution to \eqref{eq:HybridSystemGeneral}-\eqref{eq:jumpSet}. In view of the definition of the flow and jump sets, $\mathcal{C}$ and $\mathcal{D}$, in \eqref{eq:flowSet}-\eqref{eq:jumpSet}, we have that $q(0,0) \in \mathcal{C} \cup \mathcal{D}$. Suppose $q(0,0) \in \mathcal{C}\setminus\mathcal{D}$, we want to prove that $q$ is not trivial, i.e., its domain contains at least two points. For this purpose we need to show that the viability condition in \cite[Proposition 6]{heemels2021hybrid} is satisfied. 
Since the flow map $F$ is continuous and $u \in \mathcal{L_U}$, $v \in \mathcal{L_V}$ and $w \in \mathcal{L_W}$, from \cite[Proposition S1]{cortes2008discontinuous} there exists $\epsilon >0$ and an absolutely continuous function $z:[0, \epsilon] \to \mathcal{Q}$ such that $z(0) = q(0,0)$ and $\dot z(t) = F(z(t), u(t), v(t), w(t))$ for almost all $t \in [0,\epsilon]$. We write $z = (z_{x}, z_{\hat x_1}, \dots, z_{\hat x_{N+1}}, z_{\eta_1}, \dots, z_{\eta_{N+1}}, z_\sigma)$. 
Since $q(0,0) \in \mathcal{C}\setminus\mathcal{D}$, with $ \mathcal{C}\setminus\mathcal{D}$ open, and $z$ is absolutely continuous, there exists $\epsilon' \in (0,\epsilon]$ such that, for all $k \in \{1, \dots, N+1\}$, $z_{\eta_{k}}(t) \geq  z_{\eta_\sigma}(t)$ for almost all $t \in [0, \epsilon']$. Thus, $z(t) \in \mathcal{C}$ for almost all $t \in [0, \epsilon']$ and the viability condition in \cite[Proposition 6]{heemels2021hybrid} holds, which implies that $q$ is a non-trivial solution. 

To prove that $q$ is complete we need to exclude items \textit{(b)} and \textit{(c)} in \cite[Proposition 6]{heemels2021hybrid}. Item \textit{(b)} in \cite[Proposition 6]{heemels2021hybrid} occurs when at least one component of $q$ blows up in finite time, and consequently $q$ blows up in finite time. Hence, to exclude \textit{(b)} in \cite[Proposition 6]{heemels2021hybrid} we need to show that each component of $q$ must not explode in finite time. Let $q = (x, \hat x_1, \dots \hat x_{N+1}, \eta_1, \dots \eta_{N+1}, \sigma)$. From Assumption~\ref{ASS:forwardCompletenessAssumption}, $x$ cannot blow up in finite time. Moreover, $\hat x_1$ cannot do so as well in view of Theorem \ref{Prop:LyapunovSolutionProposition} and since $x$ cannot. In addition, $\hat{x}_k$, for all $k\in \{2,\dots, N+1\}$ cannot blow up in finite time in view of Lemma~\ref{Lemma:CompleteAdditionalModes} and $\eta_k$, with $k \in \{1, \dots, N+1\}$ cannot as well in view of its dynamics \eqref{eq:etaDynamics} and because $y-\hat y_k$ does not  since both $x$ and $\hat{x}_k$ do not, for all $k \in \{1 \dots, N+1\}$. Finally, $\sigma$ is constant in $\mathcal{C}$, consequently, it does not blow up in finite time. Thus, item \textit{(b)} in \cite[Proposition 6]{heemels2021hybrid} cannot occur. 
On the other hand, since $G(\mathcal{D}) \subseteq \mathcal{C} \cup \mathcal{D}$ and the jump set does not impose conditions on $u$, $v$ and $w$, item \textit{(c)} in \cite[Proposition 6]{heemels2021hybrid} cannot occur. Consequently, any maximal solution to system \eqref{eq:HybridSystemGeneral}-\eqref{eq:jumpSet} is complete. This concludes the proof. 
\hfill $\blacksquare$
%
%
%
\subsection{Average dwell-time}\label{DwellTime}
Proposition~\ref{THM:CompletenessOfMaximalSolutionsTheorem} ensures the completeness of maximal solutions under Assumptions~\ref{NominalAssumption}-\ref{ASS:forwardCompletenessAssumption}, still, Zeno phenomenon has not been ruled out yet. 
In the next proposition, we prove the existence of a uniform semiglobal average dwell-time for the solution to system \eqref{eq:HybridSystemGeneral}-\eqref{eq:jumpSet}, which thus excludes the Zeno phenomenon. Its proof is given in Appendix \ref{Proof:Prop3}. 
\begin{prop}
	Suppose Assumptions~\ref{NominalAssumption},~\ref{ASS:ass1} hold and the sets $\mathcal{V}$ and $\mathcal{W}$ are compact. Then, system \eqref{eq:HybridSystemGeneral}-\eqref{eq:jumpSet} has a \emph{uniform semiglobal average dwell-time}, i.e., for any $M \in \R_{> 0}$ there exists $c > 0$ such that any corresponding solution $q$ with $|q(0,0)|\leq M$ and $u \in \mathcal{L_U}$, $v \in \mathcal{L_V}$ and $w \in \mathcal{L_W}$, 
	is such that for any $(t,j)$, $(t',j') \in \dom q$ with $t+ j \leq t'+j'$, $ j'-j \leq \frac{1}{\tau} (t'-t) + 2$ with  $ \tau:= -\frac{1}{2\nu}\ln\left(\frac{\frac{c}{\nu}}{\varepsilon + \frac{c}{\nu}}\right)$, where $\nu$ comes from \eqref{eq:etaDynamics} and $\varepsilon$ is the design parameter in \eqref{eq:Eta_plus}. 
%
	\label{THM:AverageDwellTime}
\end{prop}

We see the importance of the parameter $\varepsilon \in \R_{> 0}$, used in the jump map for the monitoring variables \eqref{eq:Eta_plus}, in the expression of $\tau$. Indeed, if we would allow $\varepsilon$ to be equal to $0$ (which we do not), $\tau$ would have been equal to $0$. In addition, Proposition~\ref{THM:AverageDwellTime} shows that any solution $q$ to \eqref{eq:HybridSystemGeneral}-\eqref{eq:jumpSet} can exhibit at most two consecutive jumps.
Note that to obtain the results of Proposition~\ref{THM:AverageDwellTime} we do not need Assumption~\ref{ASS:forwardCompletenessAssumption}. However, in view of  Propositions~\ref{THM:CompletenessOfMaximalSolutionsTheorem} and~\ref{THM:AverageDwellTime} we have that under Assumptions~\ref{NominalAssumption}-\ref{ASS:forwardCompletenessAssumption}, for any inputs $u \in \mathcal{L_U}$, $v \in \mathcal{L_V}$,  $w \in \mathcal{L_W}$,  any maximal solution $q$ to system \eqref{eq:HybridSystemGeneral}-\eqref{eq:jumpSet} is $t$-complete, namely $\sup_t \dom q = +\infty$.
%


Now that we have established robust stability properties and the properties of the hybrid time domain of the solutions for the hybrid estimation scheme, we focus on its performance in the next section. 
\section{Performance improvement}\label{PerformanceImprovement}
The goal of this section is to establish the estimation performance improvement given by the proposed hybrid multi-observer. 
We recall that with the proposed technique we have $\eta_{\sigma(t,j)}(t,j) \leq \eta_1(t,j)$ for all $(t,j) \in \dom q$, for any solution $q$ to \eqref{eq:HybridSystemGeneral}-\eqref{eq:jumpSet} with inputs $u \in \mathcal{L_U}$, $v \in \mathcal{L_V}$ and $w \in \mathcal{L_W}$, both in the case without and with resets. Therefore, the estimation performance of the proposed hybrid multi-observer are always not worse than the performance of the nominal one according to the monitoring variables we consider.

Variable $\eta_\sigma$ defined in Section~\ref{MonitoringVariables} is a performance variable that considers the ``best" mode among the $N+1$ at any time instant: this is an instantaneous performance, which ignores the past behavior in terms of the monitoring variable. For this reason, to evaluate the performance of the proposed hybrid multi-observer, we also propose the following cost,
for any solution $q$ to \eqref{eq:HybridSystemGeneral}-\eqref{eq:jumpSet} with inputs $u \in \mathcal{L_U}$, $v \in \mathcal{L_V}$ and $w \in \mathcal{L_W}$,
for all $(t,j) \in \dom q$, 
\begin{equation}
	J_{\sigma(t,j)} (t,j) := \sum\limits_{i=0}^{j}\left(\int_{t_i}^{t_{i+1}} \eta_{\sigma(s,i)}(s,i)  \, ds\right),
	\label{eq:performanceCostSigma}
\end{equation}
with $0 = t_0 \leq t_1 \leq \dots \leq t_{j+1} = t$ satisfying $\dom q \cap ([0,t] \times \{0, 1, \dots, j\}) = \bigcup_{i=0}^j [t_i, t_{i+1}] \times \{i\}$.



Similarly, we define the performance cost of the nominal observer, for all $(t,j) \in \dom q$, as  
\begin{equation}
	J_1 (t,j) := \sum\limits_{i=0}^{j}\left(\int_{t_i}^{t_{i+1}} \eta_{1}(s,i)  \, ds\right),
	\label{eq:performanceCostNominal}
\end{equation}
with $0 = t_0 \leq t_1 \leq \dots \leq t_{j+1} = t$ satisfying $\dom q \cap ([0,t] \times \{0, 1, \dots, j\}) = \bigcup_{i=0}^j [t_i, t_{i+1}] \times \{i\}$.

In the next theorem we prove, that the hybrid scheme in Section \ref{HybridEstimationScheme} strictly improves the performance $J_1$ in \eqref{eq:performanceCostNominal}, under some conditions on the gain selection and on the initial conditions of the modes of the multi-observer and monitoring variables.
\begin{thm}
	Consider system \eqref{eq:HybridSystemGeneral}-\eqref{eq:jumpSet} under Assumptions~\ref{NominalAssumption}-\ref{ASS:forwardCompletenessAssumption}.  
	Let $q$ be a maximal solution 
	with inputs $u \in \mathcal{L_U}$,  $v \in \mathcal{L_V}$ and $w \in \mathcal{L_W}$ and for which the initial conditions of the monitoring variables are all the same, namely $\eta_k(0,0) = \eta_0$ for all $k\in \{1,\dots , N+1\}$ for some $\eta_0\in \R$. 
		Then, 
			for any $(t,j) \in \dom q$,
			\begin{equation}
				J_{\sigma(t,j)} (t,j) \leq J_1(t,j),
			\end{equation}
			with $J_\sigma$ and $J_1$ defined in \eqref{eq:performanceCostSigma} and \eqref{eq:performanceCostNominal}, respectively. 
			Moreover,
	if there exists $(t^\star, j^\star) \in \dom q$ such that 
	\begin{equation}
		\eta_{\sigma(t^\star, j^\star)}(t^\star, j^\star) < \eta_1 (t^\star, j^\star),
		\label{eq:improvementEta}
	\end{equation}
	then there exists $j^{\star'} \geq j^\star$ such that 
	\begin{equation}
		J_{\sigma(t,j)} (t,j) < J_1(t,j)
		\label{eq:performanceImprovementEquation}
	\end{equation}
	for all 
	$(t,j) \geq (t^\star, j^{\star'})$, with $(t,j) \in \dom q$. 
	\label{THM:PerformanceImprovement}
\end{thm}
\noindent\textbf{Proof:}
Consider system \eqref{eq:HybridSystemGeneral}-\eqref{eq:jumpSet} and let $q$ be a maximal solution to system \eqref{eq:HybridSystemGeneral}-\eqref{eq:jumpSet} with inputs $u \in \mathcal{L_U}$, $v \in \mathcal{L_V}$ and $w \in \mathcal{L_W}$.
From \eqref{eq:etaDynamics}, \eqref{eq:Eta_plus_nom}, \eqref{eq:Eta_plus_sigma}, \eqref{eq:flowSet}, \eqref{eq:jumpSet} and $\eta_k (0,0) = \eta_0 \in \R$ for all $k \in \{1, \dots, N+1\}$, we have, for all $(t,j) \in \dom q$,
\begin{equation}
	\eta_{\sigma(t, j)}(t, j) \leq \eta_1 (t, j). 
	\label{eq:performanceProofNotWorst}
\end{equation} 
We then derive from \eqref{eq:performanceCostSigma} and \eqref{eq:performanceCostNominal} that 
$J_{\sigma(t,j)} (t,j) \leq J_1(t,j)$, 
for all $(t,j) \in \dom q$, which concludes the first part of the proof.

In the second part of the theorem, we have that there exists $(t^\star, j^\star) \in \dom q$ such that 
\begin{equation}
	\eta_{\sigma(t^\star, j^\star)}(t^\star, j^\star) < \eta_1 (t^\star, j^\star).
	\label{eq:improvementEtaProof}
\end{equation}
We now consider two cases. If $t^\star\in \textnormal{int} \ I^{j^{\star}}$, using
\eqref{eq:performanceCostSigma}, \eqref{eq:performanceCostNominal}, \eqref{eq:performanceProofNotWorst}, \eqref{eq:improvementEtaProof}, since no jump occurs at $(t^\star,j^\star)$ and $\eta_1$ and $\eta_\sigma$ are not affected by jumps, by continuity of $\eta_\sigma$ and $\eta_1$ on $I^{j^{\star}}$, \eqref{eq:performanceImprovementEquation}
is obtained by integration of \eqref{eq:performanceProofNotWorst} for all $(t,j) \geq (t^\star, j^\star)$, with $(t,j) \in \dom q$.
On the other hand, if $I^{j^{\star}}$ is empty, since $q$ is maximal, it is $t$-complete by Propositions~\ref{THM:CompletenessOfMaximalSolutionsTheorem} and~\ref{THM:AverageDwellTime} as explained in Section \ref{DwellTime} and thus we have that there exists $j^{\star'} > j^\star$ such that $(t^\star, j^{\star'})\in \dom q$ and $\eta_{\sigma(t^\star, j^{\star'})} \leq \eta_{\sigma(t^\star, j^{\star})}  < \eta_1 (t^\star, j^{\star'})$ with $I^{j^{\star '}}$ non empty. Following similar step as before we have that \eqref{eq:performanceImprovementEquation} holds for all $(t,j) \geq (t^\star, j^{\star'})$, with $(t,j) \in \dom q$.
This concludes the proof. 
\hfill $\blacksquare$

Theorem~\ref{THM:PerformanceImprovement} shows that, if the condition in \eqref{eq:improvementEta} holds, then the cost of the proposed hybrid multi-observer $J_\sigma$ is strictly smaller than the one of the nominal observer $J_1$ and thus, the estimation performance in terms of costs $J_\sigma$ and $J_1$ is strictly improved. 

In the next proposition, we give the conditions to guarantee that \eqref{eq:improvementEta} is satisfied and consequently, from Theorem~\ref{THM:PerformanceImprovement}, that the estimation performance is strictly improved with the hybrid multi-observer \eqref{eq:HybridSystemGeneral}-\eqref{eq:jumpSet}.


\begin{prop}
	Consider system \eqref{eq:HybridSystemGeneral}-\eqref{eq:jumpSet}  with $\Lambda_2 \in \mathbb{S}^{n_x}_{> 0}$ and suppose Assumptions~\ref{NominalAssumption}-\ref{ASS:forwardCompletenessAssumption} hold.  
	Select the gains $L_k$, with $k \in \{2, \dots, N+1\}$, in \eqref{eq:observer} such that there exists $k^\star \in \{2, \dots, N+1\}$ satisfying $L_{k^\star}^\top\Lambda_2 L_{k^\star} < L_1^\top \Lambda_2 L_1$. Let $q$ be a maximal solution with inputs $u \in \mathcal{L_U}$,  $v \in \mathcal{L_V}$ and $w \in \mathcal{L_W}$ and initial condition $q(0,0)$ satisfying the following properties.
	\begin{enumerate}[label=(\roman*),leftmargin=.7cm]
		\item $\hat x_k(0,0) = \hat x_0$ for all $k \in \{1, \dots, N+1\}$ for some $\hat x_0 \in~\R^{n_x}$.
		%
		\item $\eta_k(0,0) = \eta_0$ for all  $k \in \{1, \dots, N+1\}$ for some $\eta_0\in \R$.
		\item $\hat y_k(0,0)\neq y(0,0)$ for all $k \in \{1, \dots, N+1\}$.
	\end{enumerate}
	Then, 
	there exists $(t^\star, j^\star) \in \dom q$ such that 
	\begin{equation}
		\eta_{\sigma(t^\star, j^\star)}(t^\star, j^\star) < \eta_1 (t^\star, j^\star).
	\end{equation}
	\label{THM:PerformanceImprovement_ImprovementEta}
\end{prop}
\noindent\textbf{Proof:}
Let $q$ be a maximal solution to system \eqref{eq:HybridSystemGeneral}-\eqref{eq:jumpSet} with inputs $u \in \mathcal{L_U}$, $v \in \mathcal{L_V}$ and $w \in \mathcal{L_W}$ satisfying items (\textit{i})-(\textit{iii}).
We define $\Delta_k := y -\hat{y}_k \in \R^{n_y}$ for all $k \in \{1, \dots, N+1\}$ for the sake of convenience. Note that, thanks to item (\textit{i}), 
$\Delta_1(0,0) = y(0,0) -\hat{y}_1 (0,0)
=  y(0,0) - h(\hat{x}_1 (0,0),0)
=  y(0,0) - h(\hat{x}_k (0,0),0)
=  y(0,0) - \hat{y}_k (0,0)
= \Delta_k (0,0),$
for any $k \in \{1, \dots, N+1\}$. 
On the other hand, from \eqref{eq:etaDynamics} we have, for all $k \in \{1, \dots, N+1\}$,
\begin{equation}
	\begin{aligned}
		\dot{\eta}_k &= -\nu\eta_k + (y-\hat{y}_k)^\top(\Lambda_1+ L_k^\top\Lambda_2L_k)(y-\hat{y}_k)\\ 
		&= -\nu \eta_k + \Delta_k^\top(\Lambda_1 + L_k^\top\Lambda_2 L_k)\Delta_k.
	\end{aligned}
	\label{eq:etaDotWithDelta_proofPerformanceImprovement}
\end{equation}
We evaluate \eqref{eq:etaDotWithDelta_proofPerformanceImprovement} for $k = 1$ at $(t,j) = (0,0)$. 
As $\Delta_1(0,0) = \Delta_k(0,0)$, from item (\textit{ii}) of Proposition~\ref{THM:PerformanceImprovement_ImprovementEta} and since $\Lambda_2$ is positive definite,  we obtain
\begin{equation}
	\begin{array}{l}
		\dot{\eta}_1(0,0) \\[0.5em] 
		\ = -\nu \eta_1(0,0) + \Delta_1(0,0)^\top(\Lambda_1 + L_1^\top\Lambda_2 L_1)\Delta_1(0,0)\\
		\ =-\nu \eta_{k^\star}(0,0) + \Delta_{k^\star}(0,0)^\top(\Lambda_1 + L_1^\top\Lambda_2 L_1)\Delta_{k^\star}(0,0)\\
		\ >-\nu \eta_{k^\star}(0,0) + \Delta_{k^\star}(0,0)^\top(\Lambda_1 + L_{k^\star}^\top\Lambda_2 L_{k^\star})\Delta_{k^\star}(0,0)\\
		\  = \dot{\eta}_{k^\star}(0,0)
	\end{array}
	\label{eq:etaDerivativeProofImprovement}
\end{equation}
for any $k^\star \in \{2, \dots, N+1\}$ such that $L_{k^\star}^\top\Lambda_2 L_{k^\star} < L_1^\top\Lambda_2 L_1$. The strict inequality in \eqref{eq:etaDerivativeProofImprovement} comes from the condition $L_{k^\star}^\top\Lambda_2 L_{k^\star} < L_1^\top\Lambda_2 L_1$ on the observer gain selection, with $\Lambda_2 \in \mathbb{S}^{n_x}_{> 0}$, and $\Delta_{k^\star}(0,0)\neq 0$ by item (\textit{iii}) of Proposition~\ref{THM:PerformanceImprovement_ImprovementEta}. 
Since $q$ is maximal it is $t$-complete by Propositions~\ref{THM:CompletenessOfMaximalSolutionsTheorem} and~\ref{THM:AverageDwellTime}. Moreover, $q$ can exhibit at most two consecutive jumps as explained in Section~\ref{DwellTime} and thus there exists $j^\star \in \{0,1\}$ such that $\sigma(0,j^\star) = \tilde{k}$ with $\tilde{k} \in\operatornamewithlimits{\argmin}\limits_{k \in \Pi}{\dot{\eta}_k(0,j^\star)}$, with $\Pi = \operatornamewithlimits{\argmin}\limits_{k \in \{ 1, \dots, N +1\}}{{\eta}_k(0,j^\star)}$ and $(0, j^\star +1) \notin \dom q$. Note that $L_{\tilde k}^\top\Lambda_2 L_{\tilde{k}} < L_1^\top \Lambda_2 L_1$ and $\dot q(0,j^\star) \in T_\mathcal{C}(q):= \{q \in \mathcal{Q}: \dot{\eta}_k \geq \dot{\eta}_\sigma, \forall k \in \{1, \dots, N+1\}\}$, where $\dot\eta_k=-\nu\eta_k + (y-\hat{y}_k)^\top(\Lambda_1+ L_k^\top\Lambda_2L_k)(y-\hat{y}_k)$ and $\dot\eta_\sigma=-\nu\eta_\sigma + (y-\hat{y}_\sigma)^\top(\Lambda_1+ L_\sigma^\top\Lambda_2L_\sigma)(y-\hat{y}_\sigma)$, for all $(t,j) \in \dom q$ with some abuse of notation, in view of \cite[Lemma 2.1]{petri2023thesis}.
From \eqref{eq:Eta_plus} and \eqref{eq:etaDerivativeProofImprovement}, we obtain
\begin{equation}
	\dot{\eta}_{\sigma(0,j^\star)}(0,j^\star) = \dot{\eta}_{\tilde{k}}(0,j^\star) <\dot{\eta}_1(0, j^\star).
	\label{eq:etaDerivativeZero_proofPerformanceImprovement}
\end{equation}
Moreover, since $q$ is $t$-complete, we have that there exists  $\epsilon >0$ such that $q(t,j^\star) \in \mathcal{C}$ for all $t \in [0, \epsilon]$.   
Consequently, $(t^\star, j^\star) \in \dom q^\star$ for all $t^\star \in [0, \epsilon]$. In addition, we have $\eta_{\sigma(0,j^\star)}(0,j^\star) = \eta_1(0,j^\star) = \eta_0$ both when $j^\star =0$ and $j^\star =1$ from \eqref{eq:Eta_plus_nom} and \eqref{eq:Eta_plus_sigma}. Using the last equation,  from \eqref{eq:etaDerivativeZero_proofPerformanceImprovement} 
we obtain
$	\eta_{\sigma(t^\star, j^\star)}(t^\star, j^\star) < \eta_1 (t^\star, j^\star).$
This concludes the proof.~\hfill $\blacksquare$
%

Note that, the conditions in items (\textit{i}) and (\textit{ii}) of Proposition~\ref{THM:PerformanceImprovement_ImprovementEta} can always be ensured by designing the same initial condition for the state estimate and monitoring variables for all the modes. Moreover, condition in item (\textit{iii}) is verified almost everywhere (it is a set of null measure).
%
We also acknowledge that we state the performance improvement with respect to costs $J_1$ and $J_\sigma$, and that it would be interesting to state properties for a cost, which involves the state estimation errors $e_1$ and $e_\sigma$.  This is a challenging question, which goes beyond the scope of this work.

\color{black}
\section{Numerical case studies}\label{Example}

\subsection{Van der Pol oscillator}\label{Example_VanDerPol}
We consider
\begin{equation}
	\begin{aligned}
		\dot{x} &= Ax + B\varphi(x), \quad
		y = Cx + w
	\end{aligned}
	\label{eq:VanDerPolSystem}
\end{equation}
where $x =(x_1, x_2) \in \R^2$ is the system state to be estimated, $y \in \R$ is the measured output and $w \in \R$ is the measurement noise. The system matrices are
\begin{equation}
	\begin{array}{l}
		A = \begin{bmatrix}
			0 & 1 \\
			0 & 0 \\
		\end{bmatrix}, \
		B = \begin{bmatrix}
			0 \\
			1  \\
		\end{bmatrix}, \
		C = \begin{bmatrix}
			1 & 0
		\end{bmatrix}
	\end{array}
	\label{eq:VanDerMatrices}
\end{equation}
and $\varphi(x) = \sat( - x_1 + 0.5(1-x_1^2)x_2)$ for any $x \in \R^2$, where the saturation level is symmetric and equal to $10$. We consider the measurement noise $w$ generated from a random sequence of points, occurring at intervals of $0.01$ s, with amplitude ranging between $-0.1$ and $0.1$ and with linear interpolation applied between any two consecutive points. 

We design a nominal high-gain observer for system \eqref{eq:VanDerPolSystem} 
\begin{equation}
	\begin{aligned}
		\dot{\hat{x}}_{1} &= A\hat{x}_{1} + B\varphi(\hat{x}_{1}) + L_1 (y -\hat{y}_{1}), \quad 
		\hat{y}_{1} &= C\hat{x}_{1}
	\end{aligned}
	\label{eq:VanDerPolObserver}
\end{equation}
where $\hat{x}_{1}$ is the state estimate, $\hat{y}_{1}$ is the estimated output and $L_1 \in \R^{2 \times 1}$ is the output injection gain, which is defined as $L_1 :=H_1D$, where $D \in \R^{2 \times 1}$, $H_{1} = \textnormal{diag}(h_{1}, h_{1}^2) \in \R^{2 \times 2}$, with $h_{1} \in \R_{> 0}$ the high-gain design parameter.
To satisfy Assumption~\ref{NominalAssumption}, 
$D \in \R^{2 \times 1}$ is selected such that the matrix $A-DC$ is Hurwitz and the parameter $h_{1}$ is taken sufficiently large, i.e., $h_{1} \geq h_1^\star$, where $h_1^\star$ is equal to $2\lambda_{\max}(P)K$, where $P \in \R^{2\times 2}$ is the solution of the Lyapunov equation $P(A-DC) + (A-DC)^\top P = -I_2$ and $K = 58.25$ is the Lipschitz constant of the function $\varphi$. 
We select $D$ such that the eigenvalues of $A-DC$ are equal to $-1$ and $-2$ and we obtain $D= [3, 2]$, while the parameter $h_{1}$ is selected equal to $200 > h_1^\star = 152.50$. 
With this choice of $h_{1}$, Assumption~\ref{NominalAssumption} is satisfied with a quadratic Lyapunov function and $\alpha = 53.28$. Furthermore, since the output is linear, also Assumption~\ref{ASS:ass1} is satisfied.

%
We consider $N = 4$ additional modes, with the same structure as the nominal one in \eqref{eq:VanDerPolObserver}. The only difference is the output injection gain $L_k \in \R^{2 \times 1}$, which is defined as $L_k := H_kD$, with $H_k  = \textnormal{diag}(h_{k}, h_{k}^2) \in \R^{2 \times 2}$, with $k \in \{2,\dots,5\}$. We select $h_2 = 20$, $h_3 = 1$, $h_4 =0$ and $h_5 =-1$. Note that $h_k \leq h_1^\star$, for all $k \in \{2,\dots,5\}$. Therefore,  we have no guarantees that these modes satisfy Assumption~\ref{NominalAssumption}, and consequently, that they converge. Simulations suggest that the modes with $L_2$ and $L_3$ converge, while 
the ones with $L_4$ and $L_5$ do not. 
Note that, the gain $L_4 = 0_{2\times 1}$ is the best choice to annihilate the effects of the measurement noise.


We simulate the proposed estimation technique considering the initial conditions $x(0,0) = (1,1)$, $\hat{x}_k(0,0) = (0,0)$, $\eta_k(0,0) = 10$ for all $k \in \{1,\dots, 5\}$ and $\sigma(0,0) = 1$. 
Both cases, without and with resets, are simulated 
with 
$\nu = 5$, $\Lambda_1 =1$, $\lambda_2 = 0.1\cdot I_2$ and $\varepsilon = 10^{-4}$.  Note that the condition $\nu \in (0, \alpha]$ is satisfied. 


\begin{figure}
	\centering
	\includegraphics[width=0.98\linewidth]{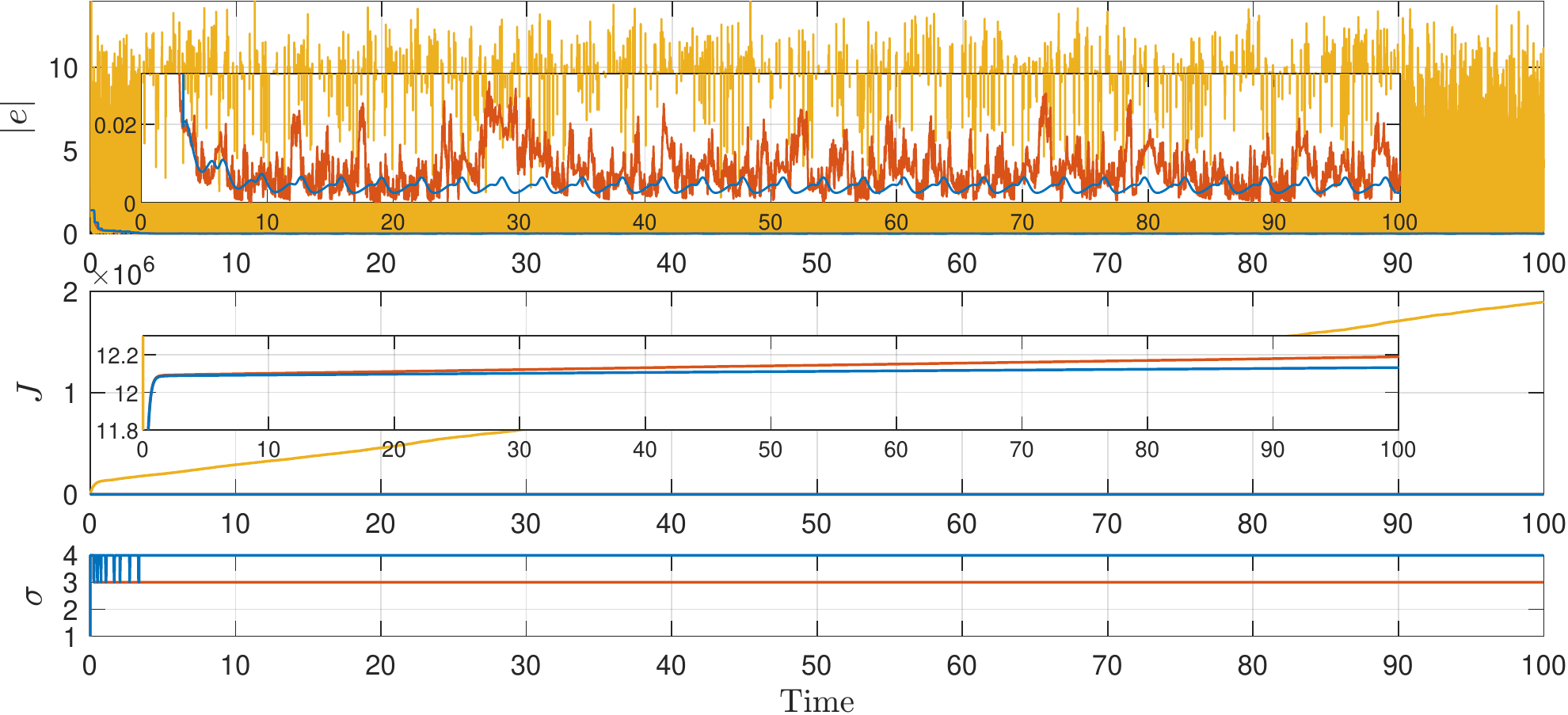}
	\caption{\normalfont{Van der Pol oscillator. Norm of the estimation error $|e|$ (top figure), performance cost $J$ (middle figure) and $\sigma$ (bottom figure). Nominal (yellow), without resets (red), with resets (blue).
	}}
	\label{FIG:estimationErrorAndSigma_VanDerPol}
\end{figure}


\setlength{\tabcolsep}{5pt}
\setlength{\extrarowheight}{1pt}
\begin{table}[]
	\centering
	\caption{Van der Pol oscillator. Average MAE and RMSE.}
	\label{Tab:MAEandRMSE}
	\begin{tabular}{l|ccc|ccc} 
		&	\multicolumn{3}{c|}{\footnotesize{no reset}}   & \multicolumn{3}{c}{\footnotesize{reset}}\\
		& $e_{1}$ & 	$e_{\sigma}$& \scriptsize{$\%$ improv. }&  $e_{1}$& 	$e_{\sigma}$ & \scriptsize{$\%$ improv.} \\
		\hline
		$\text{MAE}$&3.505 &0.031& \textbf{99.11} & 3.507 &0.035 & 98.99\\ 
		$\text{RMSE}$&3.519 &0.033& \textbf{99.05}& 3.516 & 0.036 &98.97\\ 
	\end{tabular}
\end{table}

The norm of the nominal estimation error, namely $|e_1|$, as well as $|e_\sigma|$, obtained with or without resets, 
are shown in Fig.~\ref{FIG:estimationErrorAndSigma_VanDerPol}, 
%
together with the nominal performance cost $J_1$ and 
the costs $J_{\sigma}$ obtained both in the case without and with resets. 
Fig.~\ref{FIG:estimationErrorAndSigma_VanDerPol} shows that both solutions (without resets and with resets) improve the estimation performance compared to the nominal one. 
	Zooms of the estimation error and cost plots are also shown in Fig.~\ref{FIG:estimationErrorAndSigma_VanDerPol} to highlight the difference, and thus the performance improvement, between the nominal observer and the hybrid multi-observer. 
The last plot in Fig.~\ref{FIG:estimationErrorAndSigma_VanDerPol} represents $\sigma$ and indicates which mode is selected at every time instant both in the case without and with resets. Interestingly, when the resets are considered, the fourth mode (with $L_4 = [0, 0]^\top$), that is not converging, is selected. 

To further evaluate the performance improvement given by the hybrid multi-observer, we run $100$ simulations with different initial conditions for the state estimate of all the modes of the multi-observer, both in the case without and with resets.  In particular, both components of $\hat x_k(0,0) \in \R^{2}$, for all $k \in \{1, \dots, 5\}$, were selected randomly in the interval $[-2,2]$ and, in each simulation,  all modes of the multi-observer were initialized with the same state estimate. The system state, the monitoring variables and the signal $\sigma$ were always initialized at $x(0,0) = (1,1)$, $\eta_k(0,0) = 10$, for all $k \in \{1, \dots, 5\}$, and $\sigma(0,0) = 1$. We considered the same choice of design parameters as before. 
To quantify the performance improvement,  we evaluate the mean absolute error (MAE) and the root mean square error (RMSE), averaged over all the simulations, of the state estimation error obtained with the nominal observer and the hybrid multi-observer both in the case without and with resets. The obtained data are given in Table~\ref{Tab:MAEandRMSE}. Note that the data for $e_{1}$ without and with resets are slightly different because the $100$ initial conditions were randomly selected and thus they may be different in the simulations without and with resets. Table~\ref{Tab:MAEandRMSE} shows that the proposed technique, both without and with resets, highly improves the estimation performance compared to the nominal one. Indeed, both the MAE and the RMSE are improved by more than $99\%$ both in the case without and with resets. Moreover, in this example, the performance of the hybrid multi-observer without and with resets are very similar, with the case without resets that slightly outperforms the case where the resets are implemented, both in term of MAE and RMSE. 
\subsection{Electric circuit model of a lithium-ion battery}\label{Example_Battery}
We consider an electric circuit model of $1$-cell lithium-ion battery shown in \cite[Fig. 2]{petri2021Event}, with a nonlinear output map. 
	\begin{figure}[!t]
	\centering
	\includegraphics[width=0.92\linewidth]{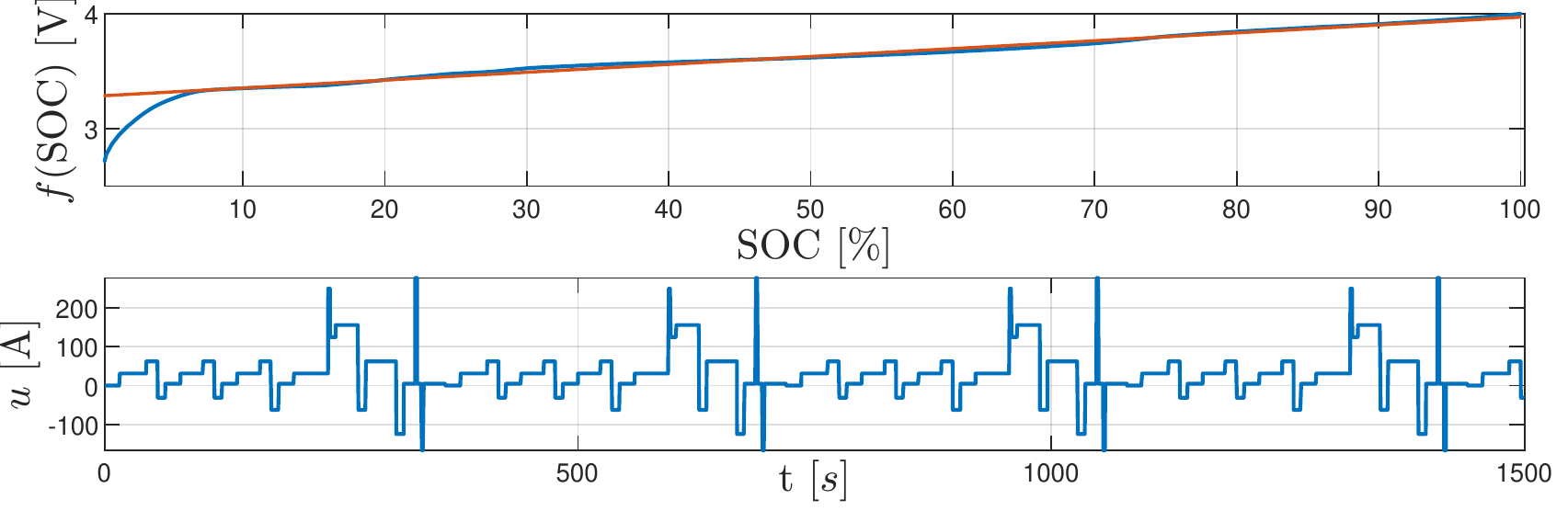}
	\caption{\normalfont{$f(\text{SOC})$(blue) with its linearization (red) and PHEV current input.}}
	\label{FIG:SOCandPHEV}
\end{figure}
From the circuit the following system model is derived
\begin{equation}
	\begin{aligned}
			\dot{x} &= Ax + Bu, \quad
			y &= Cx + f(Hx) + Du +w.
	\end{aligned}
	\label{eq:BatteryModel}
\end{equation}
The state is $x:= (U_{\text{RC}}, \text{SOC}) \in \R^2$, where $U_{\text{RC}}$ is the voltage of the RC circuit and $\text{SOC}$ is the state of charge of the battery. The output $y$ is the output voltage, the input $u$ is the current and $w$ is the measurement noise.
The system matrices are
\begin{equation}
	\begin{array}{l}
		A = \begin{bmatrix}
			-\frac{1}{\tau} & 0 \\
			0 & 0 \\
		\end{bmatrix}, \
		B = \begin{bmatrix}
			-\frac{1}{c} \\
			\frac{1}{Q}  \\
		\end{bmatrix}, \
		C = \begin{bmatrix}
			-1 & 0
		\end{bmatrix}, \\
		H = \begin{bmatrix}
		0 & 1
	\end{bmatrix}, 
	D = \begin{bmatrix}
	-R_{\text{int}}
	\end{bmatrix}. 
	\end{array}
	\label{eq:BatteryrMatrices}
\end{equation}
Considering the temperature to be constant and equal to $25^\circ \text{C}$, the parameters values are $\tau = 7 \ \text{s}$, $R = 0.5\cdot 10^{-3} \ \Omega$, $c = \frac{\tau}{R} \ \text{F}$, $Q = 25 \ \text{Ah}$ and $	R_{\text{int}} = 1 \ \text{m}\Omega$. The function $f$ and its linearization are shown in Fig.~\ref{FIG:SOCandPHEV} on the interval $[0,100] \, \%$ and we consider a first order approximation outside the interval $[0,100] \, \%$. 
%
 The function $f$ satisfies Assumption~\ref{ASS:ass1} since it has bounded derivatives. 
  The input $u$ is given by a plug-in hybrid electric vehicle (PHEV) current profile, see Fig.~\ref{FIG:SOCandPHEV}, and the measurement noise is given by $w(t) = 0.01 \sin (10 t)$, for all $t \geq 0$. 
We design the nominal observer 
 \begin{equation}
 	\begin{aligned}
 		\dot{\hat x}_1 &= A\hat x_1 + Bu + L_1(y - \hat y_1), \quad \! \!
 		\hat y_1 = C \hat x_1 + f(H\hat x_1) + Du, 
 	\end{aligned}
 	\label{eq:BatteryPolytopicObserver}
 \end{equation}
where $\hat x_1$ is the state estimate, $\hat y_1$ is the output estimate and $L_1 = [-2.07, 2.48]^\top \in \R^{2 \times 1}$ is the observer gain that is designed following a polytopic approach like in \cite{dreef2018lmi}. 
 Observer~\eqref{eq:BatteryPolytopicObserver} satisfies Assumption~\ref{NominalAssumption} with $\alpha = 0.1$. 

To improve the estimation performance, we design the hybrid multi-observer considering $N=3$ additional modes.
To select $L_2$, we linearize the output map and we design a Luenberger observer with eigenvalues in $[-0.2, -0.3]$ and we obtain $L_2 = [0.06, 61.25]^\top$. Note that, since this observer is designed for the linearized system, we have no guarantees that it satisfies an input-to-state stability property for the nonlinear system. Moreover, we chose $L_3 = [0, 0]^\top$ and we designed an extended Kalman filter \cite{reif1998ekf}, with $R_{\text{EKF}} = 1$, $Q_{\text{EKF}} = 0.1 \cdot I_2$ and $\alpha_{\text{EKF}} = 0.01$, to obtain $L_4$, which is thus a time-varying gain. Note that, in view of Remark \ref{Rem:TimeVaryingGains}, the results presented in this paper hold also in this case. 

We simulate the proposed hybrid multi-observer, both without and with resets, considering the initial conditions $x(0,0) = (1,100)$, $\hat{x}_k(0,0) = (0.5,50)$, $\eta_k(0,0) = 0$ for all $k \in \{1,\dots, 4\}$ and $\sigma(0,0) = 1$. The design parameters are selected 
$\nu = 0.05$, $\Lambda_1 =1$, $\Lambda_2 =  \begin{bsmallmatrix} 
		1 & 0 \\ 0 & 10^{-4}
	\end{bsmallmatrix}$ and $\varepsilon = 10^{-2}$.  Note that the condition $\nu \in (0, \alpha]$ is satisfied. 

\begin{figure}
	\centering
	\includegraphics[width=0.92\linewidth]{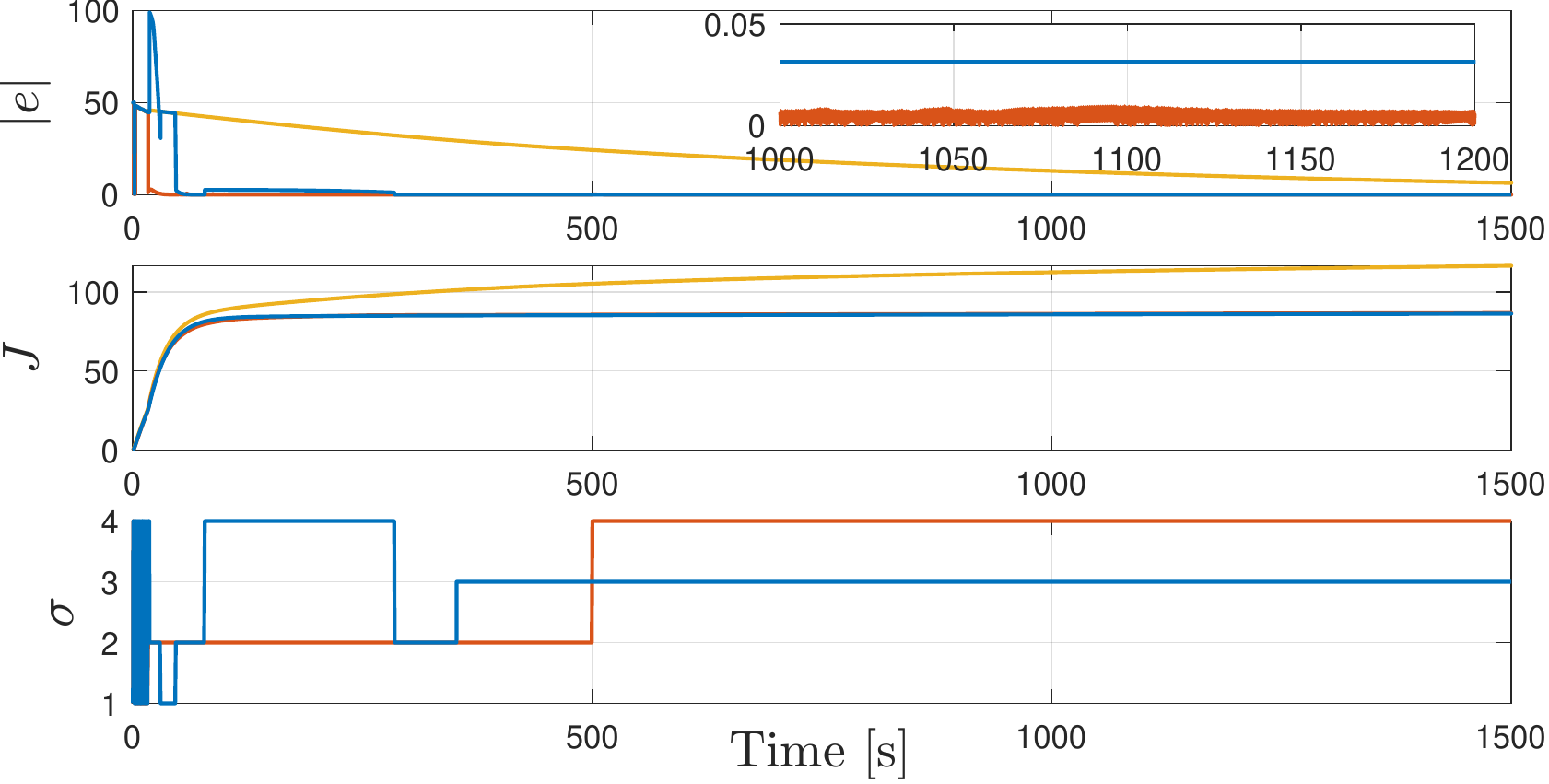}
	\caption{\normalfont{Battery example. Norm of the estimation error $|e|$ (top figure), performance cost $J$ (middle figure) and $\sigma$ (bottom figure). Nominal (yellow), without resets (red), with resets (blue).
	}}
	\label{FIG:estimationErrorAndSigma_Battery}
\end{figure}

Fig.~\ref{FIG:estimationErrorAndSigma_Battery} shows the norm of the nominal estimation error, namely $|e_1|$, as well as $|e_\sigma|$, obtained with or without resets. 
Moreover,  the nominal performance cost $J_1$ and
the costs $J_{\sigma}$ obtained both in the case without and with resets are shown in Fig.~\ref{FIG:estimationErrorAndSigma_Battery}, together with the signal $\sigma$, which indicates the selected mode at every time instant. 
Fig.~\ref{FIG:estimationErrorAndSigma_Battery} shows that both solutions (without resets and with resets) significantly improve the estimation performance compared to the nominal one. 

\setlength{\tabcolsep}{5pt}
\setlength{\extrarowheight}{1pt}
\begin{table}[]
	\centering
	\caption{Battery example. Average MAE and RMSE.} 
	\label{Tab:MAEandRMSE_Battery}
	\begin{tabular}{l|ccc|ccc}
		&	\multicolumn{3}{c|}{\footnotesize{no reset}}   & \multicolumn{3}{c}{\footnotesize{reset}}\\
		& $e_{1}$ & 	$e_{\sigma}$& \scriptsize{$\%$ improv. }&  $e_{1}$& 	$e_{\sigma}$ & \scriptsize{$\%$ improv.} \\
		\hline
		$\text{MAE}$&28.10 &3.37& 87.99 & 27.30 &1.65 & \textbf{93.94}\\ 
		$\text{RMSE}$&30.87 &8.66& 71.95& 29.90 & 5.99 &\textbf{79.98}\\ 
	\end{tabular}
\end{table}

As in the example in Section \ref{Example_VanDerPol}, we run $100$ simulations with different initial conditions for the state estimate of all modes of the multi-observer. 
In particular the first component of $\hat x_k(0,0)$ was selected randomly in the interval $[0,3] \ [\text{V}]$, while the second component of $\hat x_k(0,0)$ was selected randomly in the interval $[1,100] \ [\%]$, for all $k \in \{1, \dots, 4\}$. All the other initial conditions and the design parameters were selected as before. 
We evaluate the MAE and the RMSE as in the example in Section \ref{Example_VanDerPol} 
and the obtained results, given in Table~\ref{Tab:MAEandRMSE_Battery}, show the estimation performance improvement.
%
In this example,  the case with resets outperforms the case without resets, both in term of MAE and RMSE, according to the average results of $100$ simulations with different initial conditions showed in Table~\ref{Tab:MAEandRMSE_Battery}. However, for the specific choice of initial conditions considered to obtain the plots in Fig.~\ref{FIG:estimationErrorAndSigma_Battery} the case without resets outperforms the one with resets.

\section{Conclusion} \label{Conclusions}
We have presented a novel hybrid multi-observer that improves the state estimation performance of a given nominal nonlinear observer. Each additional mode of the multi-observer differs from the nominal one only in its output injection gain, that can be freely selected as no convergence property is required for these modes. Inspired by supervisory control/observer approaches, we have designed a switching criterion, based on monitoring variables, that selects one mode at any time instant by evaluating their performance. We have proved an input-to-state stability property of the estimation error
and the estimation performance improvement. Finally, numerical examples confirm the efficiency of the proposed approach. 

We believe that the flexibility of the presented framework leads to a range of fascinating research questions among which the off-line tuning of the additional modes gains, for which various techniques (learning based, dynamic programming) may be envisioned. Some first ideas are drawn in Section \ref{Optimization} but more could be done in this direction. The question of the link between the considered cost function, which involves the output estimation error, and a cost based on the state estimation error is also a relevant challenge to unravel.
It would also be interesting to extend the results to more general classes of systems and observers. 
This might involve investigating observability/detectability singularities or examining observers derived through immersion techniques, see e.g., \cite{rapaport2004design},
%
or observers given by differential inclusions like sliding modes observers, Levant's observers, and super-twisting observers.
Finally, merging the approach presented in this work with the one in e.g., \cite{chong2015parameter,cuevas2020multi}, addressing the case of systems with unknown parameters is also a challenging research question. 

\appendix
\subsection{Proof of Proposition~\ref{THM:LyapunovTheorem}}\label{Proof:Prop1}
%
%
The Lyapunov function of  Proposition~\ref{THM:LyapunovTheorem} is defined as 
$	U(q) := c_1 (aV(e_1) + \eta_1) + c_2 \max\limits_{k \in \{1, \dots, N+1\}}\{bV(e_k) - \eta_k, 0\} 
	+ c_3 \max\{\eta_\sigma - \eta_1, 0\},$
 for any $q \in \mathcal{Q}$, where $c_1, c_2, c_3, a, b \in \R_{> 0}$ are selected such that $c_2  <  c_3 <  c_1$,  $a > \bar{a}$ where $ \bar{a}:= \frac{\delta_1(\lambda_\text{max}(\Lambda_1) + \lambda_\text{max}(L_1^\top\Lambda_2L_1))}{\alpha} \geq 0$ and $b \in (0, \bar{b})$ with $ \bar{b}:= \frac{\lambda_\text{min}(\Lambda_1)}{\theta}$, with $\theta:= \gamma \max\limits_{k \in \{1, \dots, N+1\}}{\norm{(L_k - L_1)}^2} \in \R_{\geq 0}$, where $\gamma$ comes from Assumption \ref{NominalAssumption}. Note that $U$ is locally Lipschitz as $V$ is continuously differentiable.
%
%

We prove the three items of Proposition~\ref{THM:LyapunovTheorem} separately. 

\noindent\textit{Proof of item (i)}.
We first show the upper-bound. 
Let $q \in \mathcal{Q}$, using \eqref{eq:NominalAssumptionSandwichBound} we have
$	U(q) 
\leq c_1 (a\overline{\alpha}(|e_1|) + \eta_1) + c_2 \max\limits_{k \in  \{1, \dots, N +1\}}\{b\overline{\alpha}(|e_k|) 
-\eta_k, 0\} + c_3 \max\{\eta_\sigma - \eta_1, 0\}
\ \leq c_1 (a\overline{\alpha}(|e_1|) + \eta_1) + c_2 \sum\limits_{k =1}^{N+1}(b\overline{\alpha}(|e_k|) +  
\eta_k)  
+ c_3 ( \eta_\sigma + \eta_1)
:= \overline{\alpha}_U (|( e, \eta)|),$
for some $\overline{\alpha}_U \in \Kinf$.


We now prove the lower-bound of item (\textit{i}) of Proposition~\ref{THM:LyapunovTheorem}.  
We have that $\max\limits_{k \in  \{1, \dots, N+1\}}\{bV(e_k) - \eta_k, 0\} \geq bV(e_\sigma) - \eta_{\sigma}$ as $\sigma \in \{1, \dots, N+1\}$. 
Hence, since $\max(\eta_\sigma - \eta_1, 0) \geq \eta_\sigma - \eta_1$, in view of \eqref{eq:NominalAssumptionSandwichBound}, 
$	U(q) \geq c_1(a\underline{\alpha}(|e_1|) + \eta_1) + c_2(b\underline{\alpha}(|e_\sigma|) - \eta_\sigma)  
+ c_3 (\eta_\sigma - \eta_1)
= c_1a\underline{\alpha}(|e_1|) + (c_1-c_3)\eta_1 + c_2b\underline{\alpha}(|e_\sigma|))  
+( c_3 -c_2)\eta_\sigma.$
Since $c_1-c_3>0$ and $ c_3-c_2>0$, there exists $\underline\alpha_U\in\K_\infty$ such that 
$	U(q) \geq \underline{\alpha}_U(|(e_1, \eta_1, e_\sigma, \eta_\sigma)|)$.

\noindent\textit{Proof of item (ii)}.
For the sake of convenience we write $U(q) = U_1(q) + U_2(q) + U_3(q)$, for any $q \in \mathcal{C}$, where 
$U_1(q) = c_1 (aV(e_1) + \eta_1)$, $U_2 (q)= c_2 \max\limits_{k \in \{1, \dots, N+1\}}\{bV(e_k) - \eta_k, 0\}$ and $U_3 (q)= c_3 \max\{ \eta_\sigma - \eta_1, 0\}$. 
%
We introduce here the compact notation $F \stackrel{\Delta}{=} F(q,u,v,w)$ for the sake of convenience.
Let $q \in \mathcal{C}$, $u \in \mathcal{U}$, $v \in \mathcal{V}$ and $w \in \mathcal{W}$, in view of \eqref{eq:NominalAssumptionDerivative} and \eqref{eq:etaDynamics}, 
	$	U_1^\circ(q;F)
		 \leq - c_1a\alpha V(e_1) + c_1 a \psi_1(|v|) 
		+ c_1 a \psi_2(|w|) - c_1\nu \eta_1
	  + c_1(\lambda_\text{max}(\Lambda_1)+ \lambda_\text{max}(L_1^\top\Lambda_2L_1))|y-\hat{y}_1|^2.$
Then, using Assumption~\ref{ASS:ass1} we have $ |y -\hat y_1|^2 = |h(x,w) -h(\hat x_1,0)|^2 \leq \delta_1 V(e_1) +\delta_2|w|^2$.
%
Thus, 
	$	U_1^\circ(q;F)
		\leq
		-c_1(a \alpha - \lambda_\text{max}(\Lambda_1)\delta_1 -\lambda_\text{max}(L_1^\top\Lambda_2L_1) \delta_1)V(e_1)  
	 -c_1 \nu\eta_1 
		+c_1 a \psi_1(|v|) + c_1 a \psi_2(|w|) 
		 + c_1(\lambda_\text{max}(\Lambda_1)
		+\lambda_\text{max}(L_1^\top\Lambda_2L_1))\delta_2|w|^2.$
Since $ a  > \bar{a}= \frac{\delta_1(\lambda_\text{max}(\Lambda_1) + \lambda_\text{max}(L_1^\top\Lambda_2L_1))}{\alpha} \geq 0$ and defining $a_1:= \min\left\{\frac{a\alpha - \delta_1\lambda_\text{max}(\Lambda_1)- \delta_1 \lambda_\text{max}(L_1^\top\Lambda_2L_1)}{a}, \nu \right\}> 0$ we obtain 
\begin{equation}
	\begin{array}{l}
		U_1^\circ(q;F)
	 \leq -a_1U_1(q) + c_1 a \psi_1(|v|) + c_1 a \psi_2(|w|)  \\[0.5em]
	%
	\qquad \qquad \quad + c_1(\lambda_\text{max}(\Lambda_1) 
		  +\lambda_\text{max}(L_1^\top\Lambda_2L_1))\delta_2|w|^2.
	\end{array}
	\label{eq:U1_flow}
\end{equation}
We now consider $U_2$. We need to distinguish four cases. 

\textit{Case a)}. Suppose there exists a unique $j \in \{1, \dots, N+1\}$ such that $\max\limits_{k \in \{1, \dots, N+1\}}\{bV(e_k) - \eta_k, 0\} = bV(e_j) -\eta_j$ and $ bV(e_j) -\eta_j >0$. Then, by applying Lemma \ref{Lemma:IOSSlemma} to the $j$-th dynamics, and by recalling the definition of $\theta$ given at the beginning of the proof, we obtain
	$	U_2^\circ(q;F) 
		 =  c_2\left(b\left\langle \nabla V(e_j), f_p(x,u,v)- f_o\big(\hat{x}_j,u, L_j(y - \hat{y}_j)\big) \right\rangle \right. 
		\left.  +\nu\eta_j \right. $ $ 
		\left.
		-(y-\hat{y}_j)^\top(\Lambda_1+ L_j^\top\Lambda_2L_j)(y-\hat{y}_j)
		\right) 
		 \leq-c_2 b \alpha V(e_j) + c_2 b \psi_1(|v|) 
		+ c_2b \psi_2(|w|) 
		 + c_2b \theta |y-\hat y_j|^2 
		+ c_2 \nu \eta_j 
		-c_2(y-\hat{y}_j)^\top\Lambda_1(y-\hat{y}_j) - c_2(y-\hat{y}_j)^\top L_j^\top\Lambda_2L_j(y-\hat{y}_j)
	 \leq -c_2 (b \alpha V(e_j)-\nu\eta_j)  + c_2 b \psi_1(|v|) 
		+ c_2b \psi_2(|w|) 
	 - c_2(\lambda_\text{min}(\Lambda_1) - b\theta) |y-\hat y_j|^2.$
Since $\nu \in (0, \alpha]$ and $ b \in (0,\bar{b})$ with $\displaystyle \bar{b} = \frac{\lambda_\text{min}(\Lambda_1)}{\theta}$, $\lambda_\text{min}(\Lambda_1) - b\theta>0$ and thus, defining $a_2 := c_2\alpha$, we have
	$	U_2^\circ(q;F)
		\leq  -a_2U_2(q) + c_2 b \psi_1(|v|)  
		+ c_2b \psi_2(|w|).$

\textit{Case b)}. If for all $k \in \{1, \dots, N+1\}$, $bV(e_k)-\eta_k <0$, then $U_2(q) = 0$ and 
	$	U_2^\circ(q;F) = 0 = -a_2U_2(q).$ 

\textit{Case c)}. If there exists a subset $\mathcal{S} \subseteq \{1, \dots, N+1\}$ such that, for all $i \in \mathcal{S}$, $bV(e_i)-\eta_i =0$ and for all $j \in \{1, \dots, N+1\} \setminus \mathcal{S}$, $bV(e_j)-\eta_j <0$, then $U_2(q) = 0$. 
Following similar steps as in \textit{case a)}, we obtain
	$	U_2^\circ(q;F)= 
		 \max\limits_{i \in S}\Big\{-a_2(bV(e_i)-\eta_i) + c_2 b \psi_1(|v|) + c_2b\psi_2(|w|), 0\Big\}
		  \leq  c_2 b \psi_1(|v|)  + c_2b\psi_2(|w|)
		  = -a_2U_2(q) +  c_2 b \psi_1(|v|)  +c_2b\psi_2(|w|).$

\textit{Case d)}. If there exists a subset $\tilde{\mathcal{S}} \subseteq \{1, \dots, N+1\}$ such that, for all $i,j \in \tilde{\mathcal{S}}$, $\max\limits_{k \in \{1, \dots, N+1\}}\{bV(e_k) - \eta_k, 0\} = bV(e_i) -\eta_i = bV(e_j) - \eta_j >0$. 
Following similar steps as in \textit{case a)}, we obtain $	U_2^\circ(q;F)\leq 
 \max\limits_{i \in \tilde S}\Big\{-a_2(bV(e_i)-\eta_i) +  c_2 b \psi_1(|v|) +  c_2b\psi_2(|w|)\Big\}$.
Then, for any $i \in \tilde{S}$, by the definition of $\tilde{S}$, we have
	$	U_2^\circ(q;F) 
		\leq-a_2(bV(e_i)-\eta_i) +  c_2 b \psi_1(|v|)  +c_2b\psi_2(|w|)
		= -a_2U_2(q) +  c_2 b \psi_1(|v|) +  c_2b\psi_2(|w|).$
Merging cases \emph{a), b), c)} and \emph{d)}, 
 we have that, for any $q \in \mathcal{C}$, $u \in \mathcal{U}$, $v \in \mathcal{V}$ and $w \in \mathcal{W}$,
\begin{equation}
	\begin{array}{l}
		U_2^\circ(q;F)
	\leq -a_2U_2(q) +  c_2 b \psi_1(|v|) +  c_2b\psi_2(|w|).
	\end{array}
	\label{eq:U2_flow}
\end{equation}

We now consider $U_3$. 
Since $q \in \mathcal{C}$, from \eqref{eq:flowSet} we have that $\eta_\sigma \leq \eta_k$ for all $k \in \{1, \dots, N+1\}$. Therefore, $ \eta_\sigma \leq \eta_1$. When $ \eta_\sigma  < \eta_1$ we have that $U_3(q) = 0$ and  
\begin{equation}
	U_3^\circ(q;F)= 0 = -a_3U_3(q),
	\label{eq:U3_case1}
\end{equation}
for any $a_3 \in \R_{>0}$. When $ \eta_\sigma  = \eta_1$, since $F(q,u,v,w) \in T_\mathcal{C}(q)$ 
and $T_\mathcal{C}(q):= \{q \in \mathcal{Q}: \dot \eta_1 \geq  \dot \eta_\sigma\}$, where we use $\dot{\eta}_1 = -\nu\eta_1 + (y-\hat{y}_1)^\top(\Lambda_1+ L_1^\top\Lambda_2L_1)(y-\hat{y}_1)$ and $\dot{\eta}_\sigma = -\nu\eta_\sigma + (y-\hat{y}_\sigma)^\top(\Lambda_1+ L_\sigma^\top\Lambda_2L_\sigma)(y-\hat{y}_\sigma)$ for the sake of convenience,  we have
\begin{equation}
	U_3^\circ(q;F) \leq \max\{ \dot \eta_\sigma - \dot \eta_1, 0\} = 0 = -a_3U_3(q),
	\label{eq:U3_case2}
\end{equation}
for any $a_3 \in \R_{>0}$. Consequently, from \eqref{eq:U3_case1} and \eqref{eq:U3_case2}, we obtain that,
\begin{equation}
	U_3^\circ(q;F)\leq 0 = -a_3U_3(q),
	\label{eq:U3_flow}
\end{equation}
for all $a_3 \in \R_{>0}$. 

From \eqref{eq:U1_flow}, \eqref{eq:U2_flow} and \eqref{eq:U3_flow} we have that, for any $q \in \mathcal{C}$, $u \in \mathcal{U}$, $v \in \mathcal{V}$ and $w \in \mathcal{W}$, such that $F(q,u,v,w)\in T_\mathcal{C}(q)$,
	$	U^\circ(q;F) 
		= U_1^\circ(q,F) + 	U_2^\circ(q;F) 
		+ 	U_3^\circ(q;F)
		  \leq  -a_1U_1(q) + c_1 a \psi_1(|v|) + c_1 a \psi_2(|w|)  
		   + c_1(\lambda_\text{max}(\Lambda_1)
		+\lambda_\text{max}(L_1^\top\Lambda_2L_1))\delta_2|w|^2 
		 -a_2U_2(q) +  c_2 b \psi_1(|v|)  + c_2b\psi_2(|w|)  -a_3U_3(q). $
Defining $\alpha_0:= \min\{a_1, a_2, a_3\} \in \R_{> 0}$, 
we obtain	
	$	U^\circ(q;F)
		\leq -\alpha_0 U(q) + \phi_1(|v|) + \phi_2(|w|), $
where $\phi_1(s):= (c_1 a + c_2 b)\psi_1(s)$ and $\phi_2(s):=(c_1a +c_2b)\psi_2(s)  + (c_1(\lambda_\text{max}(\Lambda_1) 
	+\lambda_\text{max}(L_1^\top\Lambda_2L_1))\delta_2)s^2$, for any $s \geq 0$.
	
	\noindent\textit{Proof of item (iii).}
	As in the proof of item (\textit{ii}), for the sake of convenience we write $U(q) = U_1(q) + U_2(q) + U_3(q)$, for any $q \in \mathcal{D}$, where 
	$U_1(q) = c_1 (aV(e_1) + \eta_1)$, $U_2 (q)= c_2 \max\limits_{k \in \{1, \dots, N+1\}}\{bV(e_k) - \eta_k, 0\}$ and $U_3 (q)= c_3 \max\{ \eta_\sigma - \eta_1, 0\}$. 
	In addition, we use $\sigma^+$ to denote the selected mode after a jump, in view of the hybrid system notation described in the preliminaries.
	
	Let $q \in \mathcal{D}$ and $\mathfrak g \in G(q)$. Then, from \eqref{eq:StateEstimate_plus_nom} and \eqref{eq:Eta_plus_nom} we have 
	\begin{equation}
		\begin{aligned}
			U_1(\mathfrak{g}) &= c_1 (aV(e_1) + \eta_1)
			= U_1(q).
		\end{aligned}
		\label{eq:U1plus}
	\end{equation}

	We now consider $U_2$. We need to distinguish the case without resets and the case with resets. 
	We first consider the case without resets. From \eqref{eq:StateEstimate_plus_nom}-\eqref{eq:Eta_plus_compact}, we obtain 
	\begin{equation}
		\begin{aligned}
			U_2(\mathfrak{g}) &= c_2 \max\limits_{k \in \{2, \dots, N+1\}\setminus\{\sigma^+\}}\{bV(e_1) - \eta_1, bV(e_{\sigma^+}) - \eta_{\sigma^+},\\
			&\quad bV(e_k) - \eta_k -\varepsilon, 0\}\\
			&\leq c_2 \max\limits_{k \in \{2, \dots, N+1\}\setminus\{\sigma^+\}}\{bV(e_1) - \eta_1, bV(e_{\sigma^+}) - \eta_{\sigma^+},\\
			&\quad bV(e_k) - \eta_k, 0\}\\
			&= c_2 \max\limits_{k \in \{1, \dots, N+1\}}\{bV(e_k) - \eta_k, 0\}
			 = U_2(q).
		\end{aligned}
		\label{eq:U2plus_noResets}
	\end{equation}

	On the other hand, when the resets are implemented, we need to distinguish the case where $\sigma^+ = 1$ and the case when $\sigma^+ \in \{2, \dots, N+1\}$. 
	Suppose first that $\sigma^+ = 1$. Then, from \eqref{eq:StateEstimate_plus_nom}-\eqref{eq:Eta_plus_compact}, we have
	\begin{equation}
		\begin{aligned}
			U_2(\mathfrak{g}) &= c_2 \max\{bV(e_1) - \eta_1, bV(e_{k^\star}) - \eta_{\tilde k^\star} -\varepsilon, 0\}\\
			&\leq c_2 \max\{bV(e_1) - \eta_1, bV(e_{k^\star}) - \eta_{\tilde k^\star}, 0\}\\
		\end{aligned}
		\label{eq:U2plus_Resets_NomSelected_intermediateStep}
	\end{equation}
	with $k^\star \in  \operatornamewithlimits{\argmin}\limits_{k \in \Pi}(-\nu\eta_k + (y-\hat{y}_k)^\top(\Lambda_1  + L_k^\top\Lambda_2L_k)(y-\hat{y}_k))\big),$ where $\Pi(q)= \operatornamewithlimits{\argmin}\limits_{k \in \{ 1, \dots, N +1\} \setminus \{\sigma\}} \eta_k$,  
	for all $q \in \mathcal{D}$ and $\eta_{\tilde k^\star} = \operatornamewithlimits{\min}\limits_{k \in \{1, \dots, N+1\}\setminus \{\sigma\}}
	\eta_k$.
	Note that, if different observers generate the same minimum $\eta_k$, with the same minimum derivative, then $k^\star$ may be different from $\sigma^+$. However, $\eta_{\sigma^+} = \eta_{k^\star} = \tilde \eta_{k^\star}$. Consequently, from \eqref{eq:U2plus_Resets_NomSelected_intermediateStep} and since $k^\star \in \{1,\dots, N+1\}$,
	\begin{equation}
		\begin{aligned}
			U_2(\mathfrak{g})
			&\leq c_2 \max\{bV(e_1) - \eta_1, bV(e_{k^\star}) - \eta_{k^\star}, 0\}\\
			&\leq c_2 \max\limits_{k \in \{1, \dots, N+1\}}\{bV(e_k) - \eta_k, 0\}
		    = U_2(q). 
		\end{aligned}
		\label{eq:U2plus_Resets_NomSelected}
	\end{equation}
	On the contrary, when $\sigma^+ \in \{2, \dots, N+1\}$, from \eqref{eq:StateEstimate_plus_nom}-\eqref{eq:Eta_plus_compact}, we have
	\begin{equation}
		\begin{aligned}
			U_2(\mathfrak{g})
			 &= c_2 \max\{bV(e_1) - \eta_1, bV(e_{k^\star}) - \eta_{\sigma^+}, bV(e_{k^\star}) - \eta_{\tilde k^\star} \\
			 &\quad -\varepsilon, 0\}\\
			& \leq c_2 \max\{bV(e_1) - \eta_1, bV(e_{k^\star}) - \eta_{\sigma^+}, bV(e_{k^\star}) - \eta_{\tilde k^\star},\\
			 &\quad 0\},\\
		\end{aligned}
		\label{eq:U2plus_Resets_NoNomSelected_intermediateStep}
	\end{equation}
	with $k^\star \in  \operatornamewithlimits{\argmin}\limits_{k \in \Pi}(-\nu\eta_k + (y-\hat{y}_k)^\top(\Lambda_1  + L_k^\top\Lambda_2L_k)(y-\hat{y}_k))\big),$ where $\Pi(q)= \operatornamewithlimits{\argmin}\limits_{k \in \{ 1, \dots, N +1\} \setminus \{\sigma\}} \eta_k$,  for all $q \in \mathcal{D}$ and $\eta_{\tilde k^\star} = \operatornamewithlimits{\min}\limits_{k \in \{1, \dots, N+1\}\setminus \{\sigma\}}
	\eta_k$.
	Note that, if different observers generate the same minimum $\eta_k$, with the same minimum derivative, then $k^\star$ may be different from $\sigma^+$. However, $\eta_{\sigma^+} = \eta_{k^\star} = \tilde \eta_{k^\star}$. Consequently, from \eqref{eq:U2plus_Resets_NoNomSelected_intermediateStep} and since $k^\star \in \{1,\dots, N+1\}$,
		$	U_2(\mathfrak{g})
			\leq c_2 \max\{bV(e_1) - \eta_1, bV(e_{k^\star}) - \eta_{k^\star}, 0\}
			\leq c_2 \max\limits_{k \in \{1, \dots, N+1\}}\{bV(e_k) - \eta_k, 0\}
			= U_2(q). $
	As a result, from \eqref{eq:U2plus_noResets}, \eqref{eq:U2plus_Resets_NomSelected} and the last inequality 
	we have, for any $q \in \mathcal{D}$ and any $\mathfrak{g}\in G(q)$, 
	\begin{equation}
		\begin{aligned}
			U_2(\mathfrak{g}) \leq U_2(q). 
		\end{aligned}
		\label{eq:U2plus}
	\end{equation}
	
	We now consider $U_3$. From \eqref{eq:Eta_plus_nom} and \eqref{eq:Eta_plus_sigma} we have 
		$	U_3 (\mathfrak{g}) = c_3 \max\{ \eta_{\sigma^+} - \eta_1, 0\}
			 = c_3 \max\{ \eta_{\sigma} - \eta_1, 0\}
			 = U_3(q). $
	
	Merging \eqref{eq:U1plus}, \eqref{eq:U2plus} and the last equation 
	we obtain, for any $q \in \mathcal{D}$ and any $\mathfrak{g}\in G(q)$, 
			$U(\mathfrak{g}) \leq U(q), $
	which concludes the proof of item (\textit{iii}) of Proposition~\ref{THM:LyapunovTheorem}. This complete the proof.
\subsection{Proof of Proposition \ref{THM:AverageDwellTime}}\label{Proof:Prop3}
Let $u \in \mathcal{L_U}$ and $v \in \mathcal{L_V}$, $w \in \mathcal{L_W}$ with $\mathcal{V}$ and $\mathcal{W}$ compact set. Let $\overbar{M} \geq M$ such that such that $\norm{v}_\infty \leq \overbar M$ and $\norm{w}_\infty \leq \overbar M$. 
 Let $q$ be a solution to system \eqref{eq:HybridSystemGeneral} with $|q(0,0)|\leq M \leq \overbar M$. Pick any $(t,j) \in \dom q$ and let $0 = t_0 \leq t_1 \leq \dots \leq t_{j+1} = t$ satisfy $\dom q \cap ([0,t]\times \{0,1,\dots,j\}) = \bigcup_{i=0}^j [t_i, t_{i+1}] \times \{i\}$. For each $i \in \{0,\dots,j\}$ and almost all $s \in [t_i, t_{i+1}]$, $q(s,i) \in \mathcal{C}$. 
Then, from \eqref{eq:etaDynamics}, for all $k \in \{1, \dots, N+1\}$, for all $s \in (t_i, t_{i+1})$, (we omit the dependency on $(s,i)$ below),
	$	\dot \eta_k - \dot\eta_\sigma 
		 = -\nu(\eta_k - \eta_\sigma) +  (y-\hat{y}_k)^\top(\Lambda_1+ L_k^\top\Lambda_2L_k)(y-\hat{y}_k)  \: \!  -   (y-\hat{y}_\sigma)^\top(\Lambda_1+ L_\sigma^\top\Lambda_2L_\sigma)(y-\hat{y}_\sigma) 
		\geq -\nu(\eta_k - \eta_\sigma) -   (y-\hat{y}_\sigma)^\top(\Lambda_1+ L_\sigma^\top\Lambda_2L_\sigma)(y-\hat{y}_\sigma)
		\geq -\nu(\eta_k - \eta_\sigma) -  (\lambda_\text{max}(\Lambda_1) 
		+\lambda_\text{max}(L_\sigma^\top\Lambda_2L_\sigma))|y-\hat{y}_\sigma|^2. $
Then, using Assumption~\ref{ASS:ass1} we have $ |y -\hat y_\sigma|^2 = |h(x,w) -h(\hat x_\sigma,0)|^2 \leq \delta_1 V(e_\sigma) +\delta_2|w|^2$. Thus,
 from 
 we obtain, for all $s\in(t_i,t_i+1)$, 
\begin{equation}
	\begin{aligned}
		\dot \eta_k - \dot\eta_\sigma 
		&\geq -\nu(\eta_k - \eta_\sigma) -  (\lambda_\text{max}(\Lambda_1) 
			+\lambda_\text{max}(L_\sigma^\top\Lambda_2L_\sigma)) \\
		& \quad(\delta_1 V(e_\sigma) 
		+ \delta_2|w|^2). 
	\end{aligned}
	\label{eq:dwellTimeProof_2}
\end{equation}
Using \eqref{eq:LyapunovSolutionTheoremEquation}, from Theorem~\ref{Prop:LyapunovSolutionProposition}, we obtain, 
for all $(t,j) \in \dom q$, 
		$|e_\sigma(t,j)| 
		\leq \beta_U(|q(0,0)|, t) + \gamma_U(\norm{v}_\infty + \norm{w}_\infty), $
with $\beta_U \in \KL$ and $\gamma_U \in \Kinf$. 
Then, using  $|q(0,0)| \leq \overbar M$, $\norm{v}_\infty \leq \overbar M$ and $\norm{w}_\infty \leq \overbar M$ we obtain, for all $(t,j) \in \dom q$, 
	$	|e_\sigma(t,j)|
		\leq \beta_U(\overbar M, t) + \gamma_U(2 \overbar M) 
		\leq \beta_U(\overbar M, 0) + \gamma_U(2 \overbar M).$
From Assumption~\ref{NominalAssumption}, for all $e_\sigma \in \R^{n_x}$, 
	$V(e_\sigma) \leq \overline{\alpha}(|e_\sigma|), $
where $\overline{\alpha} \in \Kinf$ comes from Assumption~\ref{NominalAssumption}. 
From 
the last two inequalities
we have, for all $(t,j) \in \dom q$,
\begin{equation}
	\begin{aligned}
		V(e_\sigma(t,j)) &\leq \overline{\alpha}(\beta_U(\overbar M, t) + \gamma_U(2 \overbar  M))\\
		&\leq \check{\beta}_U(\overbar M, 0) + \check{\gamma}_U(2  \overbar M), 
	\end{aligned}
	\label{eq:dwellTimeProof_3}
\end{equation}
where $\check{\beta}_U := \overline{\alpha}\circ\beta_U \in \KL$ and $\check{\gamma}_U:= \overline{\alpha} \circ \gamma_U \in \Kinf$. 
Combining \eqref{eq:dwellTimeProof_2} with \eqref{eq:dwellTimeProof_3} we obtain, for all $k \in \{1, \dots, N+1\}$, for all $s \in [t_i, t_{i+1}]$, 
\begin{equation}
	\begin{aligned}
		\dot \eta_k - \dot\eta_\sigma 
		&\geq -\nu(\eta_k - \eta_\sigma) - (\lambda_\text{max}(\Lambda_1) 
			+\lambda_\text{max}(L_\sigma^\top\Lambda_2L_\sigma)) \\
		& \quad (\delta_1(  \check{\beta}_U(\overbar M, 0) 
		 + \check{\gamma}_U(2 \overbar M)) + \delta_2 \overbar M^2)\\ 
		&\geq -\nu(\eta_k - \eta_\sigma) - c,
	\end{aligned}
	\label{eq:dwellTimeProof_4}
\end{equation}
with $c:= (\lambda_\text{max}(\Lambda_1)+ \operatornamewithlimits{\max}\limits_{k \in \{1, \dots, N +1\}}\lambda_\text{max}(L_k^\top\Lambda_2L_k)) (\delta_1(  \check{\beta}_U($ $ \overbar M, 0)  + \check{\gamma}_U(2  \overbar M)) + \delta_2 \overbar M^2) \in \R_{> 0}$. 
Integrating \eqref{eq:dwellTimeProof_4} and applying the comparison principle \cite[Lemma 3.4]{khalil2002nonlinear} we obtain, for all $s \in [t_i,t_{i+1}]$, for all $k \in \{1, \dots, N+1\}$, 
\begin{equation}
	\begin{aligned}
		\eta_k(s,i)-\eta_\sigma(s,i) &\geq e^{-\nu(s-t_i)}(\eta_k(t_i,i)-\eta_\sigma(t_i,i))\\
		&\quad - \frac{c}{\nu}(1-e^{-\nu(s-t_i)}).\\
	\end{aligned}
	\label{eq:dwellTimeProof_5}
\end{equation}	
On the other hand, from \eqref{eq:jumpSet}, we have
\begin{equation}
	t_{i+1}:= \inf\{t \geq t_i: \operatornamewithlimits{\min}\limits_{k \in \{1, \dots, N +1\}\setminus \{\sigma\}}\eta_k(t,i) = \eta_\sigma(t,i)\}. 
	\label{eq:dwellTimeProof_nextJump}
\end{equation}
We define
 $k^\star := \sigma(t_{i+1}, i+1) \in  \operatornamewithlimits{\argmin}\limits_{k \in \Pi}(-\nu\eta_k(t_{i+1}, i+1) + (y(t_{i+1}, i+1)-\hat{y}_k(t_{i+1}, i+1))^\top(\Lambda_1  + L_k^\top\Lambda_2L_k)(y(t_{i+1}, i+1)-\hat{y}_k(t_{i+1}, i+1)))\big),$ where $\Pi(q)= \operatornamewithlimits{\argmin}\limits_{k \in \{ 1, \dots, N +1\} \setminus \{\sigma\}} \eta_k$.
%
Evaluating \eqref{eq:dwellTimeProof_5} for $s = t_{i+1}$ and $k = k^\star$, from \eqref{eq:dwellTimeProof_nextJump}, we have
\begin{equation}
	\begin{aligned}
		0 &= \eta_{k^\star}(t_{i+1},i) -\eta_\sigma(t_{i+1},i)	\\
		&\geq e^{-\nu(t_{i+1}-t_i)}(\eta_{k^\star}(t_i,i)-\eta_\sigma(t_i,i)) - \frac{c}{\nu}(1-e^{-\nu(t_{i+1}-t_i)}).
	\end{aligned}
	\label{eq:dwellTimeProof_bound}
\end{equation}

We first consider the case where $k^\star \neq 1$. Note that $\sigma(s, i) \neq k^\star$ by the definition of $k^\star$, for all $s \in [t_i, t_{i+1}]$. We now consider the cases without and with resets separately. From \eqref{eq:Eta_plus_sigma}, \eqref{eq:Eta_plus} and \eqref{eq:jumpSet} we have in the case without resets, for all $i \in \Z_{>0}$, 
$	\eta_{k^\star} (t_i,i) = \eta_{k^\star} (t_i,i-1) + \varepsilon \geq \eta_\sigma (t_i,i) + \varepsilon,$
while in the case with resets, 
	$\eta_{k^\star} (t_i,i) = \eta_\sigma (t_i,i) + \varepsilon.$
As a result,
$	\eta_{k^\star} (t_i,i) \geq \eta_\sigma (t_i,i) + \varepsilon, $
both in the case without and with resets. 
Thus, $\eta_{k^\star}(t_i,i)-\eta_\sigma(t_i,i) \geq \varepsilon$ and from \eqref{eq:dwellTimeProof_bound}, 
	$0 \geq  e^{-\nu(t_{i+1}-t_i)}\varepsilon - \frac{c}{\nu}(1-e^{-\nu(t_{i+1}-t_i)}), $
which can be rewritten as
$ e^{-\nu(t_{i+1}-t_i)}\left(\varepsilon + \frac{c}{\nu}\right) \leq \frac{c}{\nu}$,
that implies
\begin{equation}
	t_{i+1}-t_i \geq -\frac{1}{\nu}\ln\left(\frac{\frac{c}{\nu}}{\varepsilon + \frac{c}{\nu}}\right) \in \R_{> 0}. 
	\label{eq:dwellTimeProof_sigmaNeq1_2}
\end{equation}

On the other hand, when $k^\star = 1 = \sigma(s, i+1)$, for all $s \in [t_{i+1}, t_{i+2}]$, we have that $\sigma(t_{i+2}, i+2) = \operatornamewithlimits{\argmin}\limits_{k \in \{ 2, \dots, N +1\}} \eta_k(t_{i+1}, i)) \neq 1$. Therefore, from \eqref{eq:dwellTimeProof_sigmaNeq1_2}, we obtain $ t_{i+2}-t_{i+1} \geq -\frac{1}{\nu}\ln\left(\frac{\frac{c}{\nu}}{\varepsilon + \frac{c}{\nu}}\right) \in \R_{> 0}$. 
Consequently, for all switching times $(t_i,i) \in \dom q$, we have
$	t_{i+2}-t_i \geq -\frac{1}{\nu}\ln\left(\frac{\frac{c}{\nu}}{\varepsilon + \frac{c}{\nu}}\right) \in \R_{> 0}.$
%
Pick any $(t,j), (t',j') \in \dom q$ such that $t+j \leq t'+j'$, from the last inequality
and using $ \tau= -\frac{1}{2\nu}\ln\left(\frac{\frac{c}{\nu}}{\varepsilon + \frac{c}{\nu}}\right)$ we obtain 
$\displaystyle j'-j \leq \frac{1}{\tau} (t'-t) + 2$, 
which concludes the proof. 	

\bibliography{bibliography}
\begin{IEEEbiography}
	{Elena Petri} is a postdoc researcher at the Eindhoven University of Technology, The Netherlands.
	She obtained the Ph.D. degree in Control Engineering from the Universit\'e de Lorraine, France in 2023 and the bachelor and master degrees in Mechatronics Engineering from the University of Padova, Italy, in 2017 and 2020, respectively. 
	Her research interests include observers, nonlinear systems and hybrid systems.
\end{IEEEbiography}
\begin{IEEEbiography}
	{Romain Postoyan} received the “Ingénieur” degree in Electrical and Control Engineering from ENSEEIHT (France) in 2005. He obtained the M.Sc. by Research in Control Theory \& Application from Coventry University (United Kingdom) in 2006 and the Ph.D. in Control Theory from Universit\'e Paris-Sud (France) in 2009. In 2010, he was a research assistant at the University of Melbourne (Australia). Since 2011, he is a CNRS researcher at the CRAN (France).
\end{IEEEbiography}
\begin{IEEEbiography}
	 %
	 {Daniele Astolfi} obtained a joint Ph.D. degree in Control Theory from the University of Bologna (Italy) and from Mines ParisTech (France), in 2016.
	 Afterwards he spent  two years at CRAN, Nancy (France) as a postdoc researcher.
	 Since 2018, he is a CNRS Researcher at LAGEPP, Lyon, France. He was a recipient of the 2016 Best Italian Ph.D. Thesis Award given by SIDRA and nominated for the best student paper award at ECC16 and best paper award at NOLCOS2019. He serves as Associate Editor for Automatica since 2018.
\end{IEEEbiography}
\begin{IEEEbiography}
	{Dragan  Ne\v{s}i\'c} is a Professor at The University of Melbourne, Australia. 
	His research interests include networked control systems, reset systems, extremum seeking control, hybrid control systems, event-triggered control, security and privacy in cyber-physical systems, and so on. He is a Fellow of the Institute of Electrical and Electronic Engineers (IEEE, 2006) and Fellow of the International Federation for Automatic Control (IFAC, 2019). He was a co-recipient of the George S. Axelby Outstanding Paper Award for the Best Paper in IEEE Transactions on Automatic Control (2018). 
\end{IEEEbiography}
\begin{IEEEbiography}
	{Vincent Andrieu} is a Senior Reseacher at CNRS (Directeur de recherche). He graduated in applied mathematics from INSA de Rouen, France, in 2001. After working in ONERA (French aerospace research company), he obtained a PhD degree in control theory from Ecole des Mines de Paris in 2005. In 2006, he had a research appointment at the Control and Power Group, Dept. EEE, Imperial College London. In 2008, he joined the CNRSLAAS lab in Toulouse, France, as a CNRS-charge de recherche. Since 2010, he has been working in LAGEPP-CNRS, Universite de Lyon 1, France. In 2014, he joined the functional analysis group from Bergische Universitat Wuppertal in Germany, for two sabbatical years.
\end{IEEEbiography}
\end{document}

%% file: commands.tex
\newcommand{\R}{\ensuremath{\mathbb{R}}}

\newcommand{\Rlo}{\ensuremath{\mathbb{R}_{\geq 0}}}

\newcommand{\Zo}{\ensuremath{\mathbb{Z}_{\geq 0}}}
\newcommand{\Zp}{\ensuremath{\mathbb{Z}_{> 0}}}
\newcommand{\Z}{\ensuremath{\mathbb{Z}}}

\definecolor{bleucit}{rgb}{0.2,0.4,0.6} 

\definecolor{blue_cv}{rgb}{0.09,0.35,0.78}

\newcommand{\KL}{\ensuremath{\mathcal{KL}}}
\newcommand{\K}{\ensuremath{\mathcal{K}}}
\newcommand{\Kinf}{\ensuremath{\mathcal{K}_{\infty}}}

\newcommand{\argmin}{\ensuremath{\text{argmin}\,}}

\newcommand{\dom}{\ensuremath{\text{dom}\,}}

\newcommand{\sat}{\ensuremath{\text{sat}}}

\newcommand{\norm}[1]{\ensuremath{\left\|{#1}\right\|}}

\theoremstyle{theorem}

\newtheorem{ass}{\textnormal{\textbf{Assumption}}}
\newtheorem{prop}{Proposition}

\newtheorem{lem}{Lemma}

\newtheorem{thm}{\textnormal{\textbf{Theorem}}}

\newtheorem{rem}{\textnormal{\textbf{Remark}}}

%% file: HybridMultiObserverSecondVersion.bbl
\begin{thebibliography}{10}

\bibitem{bernard2022observer}
P.~Bernard, V.~Andrieu, and D.~Astolfi, ``Observer design for continuous-time
  dynamical systems,'' {\em Annual Reviews in Control}, vol.~53, pp.~224--248,
  2022.

\bibitem{kalman1961new}
R.~E. Kalman and R.~S. Bucy, ``New results in linear filtering and prediction
  theory,'' {\em Journal of Basic Engineering}, vol.~83, no.~1, pp.~95--108,
  1961.

\bibitem{helton1999extending}
J.~W. Helton and M.~R. James, {\em Extending $\mathcal{H}_\infty$ Control to
  Nonlinear Systems: Control of Nonlinear Systems to Achieve Performance
  Objectives}.
\newblock SIAM, 1999.

\bibitem{li2015finite}
Y.~Li and R.~G. Sanfelice, ``A finite-time convergent observer with robustness
  to piecewise-constant measurement noise,'' {\em Automatica}, vol.~57,
  pp.~222--230, 2015.

\bibitem{rios2018hybrid}
H.~R{\'\i}os and A.~R. Teel, ``A hybrid fixed-time observer for state
  estimation of linear systems,'' {\em Automatica}, vol.~87, pp.~103--112,
  2018.

\bibitem{alessandri2022hysteresis}
A.~Alessandri and R.~G. Sanfelice, ``Hysteresis-based switching observers for
  linear systems using quadratic boundedness,'' {\em Automatica}, vol.~136,
  p.~109982, 2022.

\bibitem{popov2017adaptive}
I.~Popov, P.~Koschorrek, A.~Haghani, and T.~Jeinsch, ``Adaptive {K}alman
  filtering for dynamic positioning of marine vessels,'' {\em
  IFAC-PapersOnLine}, vol.~50, no.~1, pp.~1121--1126, 2017.

\bibitem{astolfi2015high}
D.~Astolfi and L.~Marconi, ``A high-gain nonlinear observer with limited gain
  power,'' {\em IEEE Transactions on Automatic Control}, vol.~60, no.~11,
  pp.~3059--3064, 2015.

\bibitem{astolfi2018low}
D.~Astolfi, L.~Marconi, L.~Praly, and A.~R. Teel, ``Low-power peaking-free
  high-gain observers,'' {\em Automatica}, vol.~98, pp.~169--179, 2018.

\bibitem{esfandiari2019bank}
K.~Esfandiari and M.~Shakarami, ``Bank of high-gain observers in output
  feedback control: Robustness analysis against measurement noise,'' {\em IEEE
  Transactions on Systems, Man, and Cybernetics: Systems}, vol.~51, no.~4,
  pp.~2476--2487, 2019.

\bibitem{bernat2015multi}
J.~Bernat and S.~Stepien, ``Multi-modelling as new estimation schema for
  high-gain observers,'' {\em International Journal of Control}, vol.~88,
  no.~6, pp.~1209--1222, 2015.

\bibitem{ahrens2009high}
J.~H. Ahrens and H.~K. Khalil, ``High-gain observers in the presence of
  measurement noise: A switched-gain approach,'' {\em Automatica}, vol.~45,
  no.~4, pp.~936--943, 2009.

\bibitem{farza2021improved}
M.~Farza, A.~Ragoubi, S.~Hadj~Sa{\"\i}d, and M.~M’Saad, ``Improved high gain
  observer design for a class of disturbed nonlinear systems,'' {\em Nonlinear
  Dynamics}, vol.~106, pp.~631--655, 2021.

\bibitem{reif1998ekf}
K.~Reif, F.~Sonnemann, and R.~Unbehauen, ``An {EKF}-based nonlinear observer
  with a prescribed degree of stability,'' {\em Automatica}, vol.~34, no.~9,
  pp.~1119--1123, 1998.

\bibitem{bonnabel2014contraction}
S.~Bonnabel and J.-J. Slotine, ``A contraction theory-based analysis of the
  stability of the deterministic extended kalman filter,'' {\em IEEE
  Transactions on Automatic Control}, vol.~60, no.~2, pp.~565--569, 2014.

\bibitem{dai2022q}
X.~Dai, V.~Nateghi, H.~Fourati, and C.~Prieur, ``Q-learning-based noise
  covariance adaptation in kalman filter for marg sensors attitude
  estimation,'' in {\em IEEE International Symposium on Inertial Sensors and
  Systems, \textnormal{Avignon, France}}, pp.~1--6, 2022.

\bibitem{astolfi2021stubborn}
D.~Astolfi, A.~Alessandri, and L.~Zaccarian, ``Stubborn and dead-zone redesign
  for nonlinear observers and filters,'' {\em IEEE Transactions on Automatic
  Control}, vol.~66, no.~2, pp.~667--682, 2021.

\bibitem{battilotti2021performance}
S.~Battilotti, ``Performance optimization via sequential processing for
  nonlinear state estimation of noisy systems,'' {\em IEEE Transactions on
  Automatic Control}, vol.~67, no.~2, pp.~2957--2972, 2021.

\bibitem{astolfi2021constrained}
D.~Astolfi, P.~Bernard, R.~Postoyan, and L.~Marconi, ``Constrained state
  estimation for nonlinear systems: a redesign approach based on convexity,''
  {\em IEEE Transactions on Automatic Control}, vol.~67, no.~2, pp.~824--839,
  2022.

\bibitem{astolfi2019uniting}
D.~Astolfi, R.~Postoyan, and D.~Ne{\v{s}}i{\'c}, ``Uniting observers,'' {\em
  IEEE Transactions on Automatic Control}, vol.~65, no.~7, pp.~2867--2882,
  2020.

\bibitem{sarkka2007unscented}
S.~Sarkka, ``On unscented kalman filtering for state estimation of
  continuous-time nonlinear systems,'' {\em IEEE Transactions on Automatic
  Control}, vol.~52, no.~9, pp.~1631--1641, 2007.

\bibitem{schiller2022suboptimal}
J.~D. Schiller and M.~A. M{\"u}ller, ``Suboptimal nonlinear moving horizon
  estimation,'' {\em IEEE Transactions on Automatic Control}, vol.~68, no.~4,
  pp.~2199--2214, 2022.

\bibitem{schiller2022simple}
J.~D. Schiller, B.~Wu, and M.~A. M{\"u}ller, ``A simple suboptimal moving
  horizon estimation scheme with guaranteed robust stability,'' {\em IEEE
  Control Systems Letters}, vol.~7, pp.~19--24, 2022.

\bibitem{peralez2022neural}
J.~Peralez, M.~Nadri, and D.~Astolfi, ``Neural network-based {KKL} observer for
  nonlinear discrete-time systems,'' in {\em IEEE Conference on Decision and
  Control, \textnormal{Canc{\'u}n, Mexico}}, pp.~2105--2110, 2022.

\bibitem{chong2015parameter}
M.~S. Chong, D.~Ne{\v{s}}i{\'c}, R.~Postoyan, and L.~Kuhlmann, ``Parameter and
  state estimation of nonlinear systems using a multi-observer under the
  supervisory framework,'' {\em IEEE Transactions on Automatic Control},
  vol.~60, no.~9, pp.~2336--2349, 2015.

\bibitem{aguiar2008identification}
A.~P. Aguiar, V.~Hassani, A.~M. Pascoal, and M.~Athans, ``Identification and
  convergence analysis of a class of continuous-time multiple-model adaptive
  estimators,'' {\em IFAC Proceedings Volumes}, vol.~41, no.~2, pp.~8605--8610,
  2008.

\bibitem{aguiar2007convergence}
A.~P. Aguiar, M.~Athans, and A.~M. Pascoal, ``Convergence properties of a
  continuous-time multiple-model adaptive estimator,'' {\em European Control
  Conference, \textnormal{Kos, Greece}}, pp.~1530--1536, 2007.

\bibitem{meijer2021joint}
T.~J. Meijer, V.~S. Dolk, M.~S. Chong, R.~Postoyan, B.~de~Jager,
  D.~Ne{\v{s}}i{\'c}, and W.~P. M.~H. Heemels, ``Joint parameter and state
  estimation of noisy discrete-time nonlinear systems: A supervisory
  multi-observer approach,'' in {\em IEEE Conference on Decision and Control,
  \textnormal{Austin, USA}}, pp.~5163--5168, 2021.

\bibitem{cuevas2020multi}
L.~Cuevas, D.~Ne{\v{s}}i{\'c}, C.~Manzie, and R.~Postoyan, ``A multi-observer
  approach for parameter and state estimation of nonlinear systems with slowly
  varying parameters,'' {\em IFAC-PapersOnLine}, vol.~53, no.~2,
  pp.~4208--4213, 2020.

\bibitem{han2018simple}
W.~Han, H.~L. Trentelman, Z.~Wang, and Y.~Shen, ``A simple approach to
  distributed observer design for linear systems,'' {\em IEEE Transactions on
  Automatic Control}, vol.~64, no.~1, pp.~329--336, 2019.

\bibitem{park2016design}
S.~Park and N.~C. Martins, ``Design of distributed {LTI} observers for state
  omniscience,'' {\em IEEE Transactions on Automatic Control}, vol.~62, no.~2,
  pp.~561--576, 2017.

\bibitem{wang2022hybrid}
L.~Wang, J.~Liu, and A.~S. Morse, ``A hybrid observer for estimating the state
  of a distributed linear system,'' {\em Automatica}, vol.~146, p.~110633,
  2022.

\bibitem{shim2015nonlinear}
H.~Shim and D.~Liberzon, ``Nonlinear observers robust to measurement
  disturbances in an {ISS} sense,'' {\em IEEE Transactions on Automatic
  Control}, vol.~61, no.~1, pp.~48--61, 2016.

\bibitem{hespanha2003hysteresis}
J.~P. Hespanha, D.~Liberzon, and A.~S. Morse, ``Hysteresis-based switching
  algorithms for supervisory control of uncertain systems,'' {\em Automatica},
  vol.~39, no.~2, pp.~263--272, 2003.

\bibitem{morse1996supervisory}
A.~S. Morse, ``Supervisory control of families of linear set-point
  controllers-part i. exact matching,'' {\em IEEE Transactions on Automatic
  Control}, vol.~41, no.~10, pp.~1413--1431, 1996.

\bibitem{hespanha2001multiple}
J.~P. Hespanha, D.~Liberzon, A.~S. Morse, B.~D.~O. Anderson, T.~S. Brinsmead,
  and F.~De~Bruyne, ``Multiple model adaptive control. part 2: switching,''
  {\em International Journal of Robust and Nonlinear Control}, vol.~11, no.~5,
  pp.~479--496, 2001.

\bibitem{peralez2020data}
J.~Peralez, F.~Galuppo, P.~Dufour, C.~Wolf, and M.~Nadri, ``Data-driven
  multi-model control for a waste heat recovery system,'' in {\em IEEE
  Conference on Decision and Control, \textnormal{Jeju Island, South Korea}},
  pp.~5501--5506, 2020.

\bibitem{goebel2012hybrid}
R.~Goebel, R.~G. Sanfelice, and A.~R. Teel, {\em Hybrid Dynamical Systems:
  Modeling, Stability, and Robustness}.
\newblock New Jersey, USA, Princeton University Press, 2012.

\bibitem{petri2023state}
E.~Petri, T.~Reynaudo, R.~Postoyan, D.~Astolfi, D.~Ne{\v{s}}i{\'c}, and
  S.~Ra\"{e}l, ``State estimation of an electrochemical lithium-ion battery
  model: improved observer performance by hybrid redesign,'' {\em European
  Control Conference, \textnormal{Bucharest, Romania}}, pp.~2151--2156, 2023.

\bibitem{petri2022towards}
E.~Petri, R.~Postoyan, D.~Astolfi, D.~Ne{\v{s}}i{\'c}, and V.~Andrieu,
  ``Towards improving the estimation performance of a given nonlinear observer:
  a multi-observer approach,'' {\em IEEE Conference on Decision and Control,
  \textnormal{Canc{\'u}n, Mexico}}, pp.~583--590, 2022.

\bibitem{heemels2021hybrid}
W.~P. M.~H. Heemels, P.~Bernard, K.~J.~A. Scheres, R.~Postoyan, and R.~G.
  Sanfelice, ``Hybrid systems with continuous-time inputs: Subtleties in
  solution concepts and existence properties,'' {\em IEEE Conference on
  Decision and Control, \textnormal{Austin, USA}}, pp.~5361--5366, 2021.

\bibitem{clarke1990optimization}
F.~H. Clarke, {\em Optimization and Nonsmooth Analysis}.
\newblock Philadelphia, USA, Classic in Applied Mathematics vol. 5, SIAM, 1990.

\bibitem{seron2012fundamental}
M.~M. Seron, J.~H. Braslavsky, and G.~C. Goodwin, {\em Fundamental limitations
  in filtering and control}.
\newblock Springer Science \& Business Media, 2012.

\bibitem{willems2004deterministic}
J.~C. Willems, ``Deterministic least squares filtering,'' {\em Journal of
  Econometrics}, vol.~118, no.~1-2, pp.~341--373, 2004.

\bibitem{na2017adaptive}
J.~Na, G.~Herrmann, and K.~G. Vamvoudakis, ``Adaptive optimal observer design
  via approximate dynamic programming,'' {\em American Control Conference,
  \textnormal{Seattle, USA}}, pp.~3288--3293, 2017.

\bibitem{petri2023thesis}
E.~Petri, ``Hybrid techniques for state estimation: event-triggered sampling
  and performance improvement,'' {\em PhD thesis, Universit{\'e} de Lorraine},
  2023.

\bibitem{khalil2014high}
H.~K. Khalil and L.~Praly, ``High-gain observers in nonlinear feedback
  control,'' {\em International Journal of Robust and Nonlinear Control},
  vol.~24, no.~6, pp.~993--1015, 2014.

\bibitem{cai2007smooth}
C.~Cai, A.~R. Teel, and R.~Goebel, ``Smooth {L}yapunov functions for hybrid
  systems-part {I}: Existence is equivalent to robustness,'' {\em IEEE
  Transactions on Automatic Control}, vol.~52, no.~7, pp.~1264--1277, 2007.

\bibitem{sontag2008input}
E.~D. Sontag, ``Input to state stability: Basic concepts and results,'' in {\em
  Nonlinear and {O}ptimal {C}ontrol {T}heory}, pp.~163--220, Springer, 2008.

\bibitem{khalil2002nonlinear}
H.~K. Khalil, {\em Nonlinear Systems}, vol.~3.
\newblock Prentice hall Upper Saddle River, NJ, 2002.

\bibitem{cortes2008discontinuous}
J.~Cortes, ``Discontinuous dynamical systems,'' {\em IEEE Control systems
  magazine}, vol.~28, no.~3, pp.~36--73, 2008.

\bibitem{petri2021Event}
E.~Petri, R.~Postoyan, D.~Astolfi, D.~Ne{\v{s}}i{\'c}, and W.~P. M.~H. Heemels,
  ``Event-triggered observer design for linear systems,'' {\em IEEE Conference
  on Decision and Control, \textnormal{Austin, USA}}, pp.~546--551, 2021.

\bibitem{dreef2018lmi}
H.~Dreef, H.~Beelen, and M.~Donkers, ``{LMI}-based robust observer design for
  battery state-of-charge estimation,'' in {\em IEEE Conference on Decision and
  Control, \textnormal{Miami, USA}}, pp.~5716--5721, 2018.

\bibitem{rapaport2004design}
A.~Rapaport and A.~Maloum, ``Design of exponential observers for nonlinear
  systems by embedding,'' {\em International Journal of Robust and Nonlinear
  Control}, vol.~14, no.~3, pp.~273--288, 2004.

\end{thebibliography}
